\documentclass[aps,prd,twocolumn,amsmath,superscriptaddress,amssymb,showpacs,floatfix,nofootinbib,longbibliography]{revtex4-1}
\usepackage{graphicx}
\usepackage{dcolumn}
\usepackage{xcolor}
\usepackage{bm}
\usepackage{natbib}
\usepackage{calligra}
\usepackage[T1]{fontenc}
\usepackage{egothic}
\usepackage[T1]{fontenc}
\newfont{\rsfsten}{rsfs10 scaled 1200}
\newfont{\rsfsseven}{rsfs10 scaled 1200}
\newfont{\rsfsfive}{rsfs10 scaled 1200}
\usepackage{epsfig}
\usepackage{units}
\usepackage[utf8]{inputenc}
\usepackage{hyperref}

\newcommand{\aap}{{Astron.~Astrophys.}}

\newcommand{\ssr}{{Space~Science~Rev.}}

\newcommand{\be}{\begin{equation}}
\newcommand{\ee}{\end{equation}}
\newcommand{\bea}{\begin{eqnarray}}
\newcommand{\eea}{\end{eqnarray}}

\def\lsim{\mathrel{\raise.3ex\hbox{$<$\kern-.75em\lower1ex\hbox{$\sim$}}}}
\def\gsim{\mathrel{\raise.3ex\hbox{$>$\kern-.75em\lower1ex\hbox{$\sim$}}}}

\begin{document}

\title{Constraining the Charge-, Time- and Rigidity-Dependence of Cosmic-Ray Solar Modulation with AMS-02 Observations during Solar Cycle 24}

\author{Ilias Cholis}
\email{cholis@oakland.edu, ORCID: orcid.org/0000-0002-3805-6478}
\affiliation{Department of Physics, Oakland University, Rochester, Michigan, 48309, USA}
\author{Ian McKinnon}
\email{ianmckinnon@oakland.edu}
\affiliation{Department of Physics, Oakland University, Rochester, Michigan, 48309, USA}
\date{\today}

\begin{abstract}
Our basic theoretical understanding of the sources of cosmic rays and their propagation through the interstellar 
medium is hindered by the Sun, that through the solar wind affects the observed cosmic-ray spectra. This effect 
is known as solar modulation. Recently released cosmic-ray observations from the Alpha Magnetic Spectrometer 
(\textit{AMS-02}) and publicly available measurements of the solar wind properties from the Advanced Composition 
Explorer and the Wilcox observatory allow us to test the analytical modeling of the time-, charge- and rigidity-dependence 
of solar modulation. We rely on associating measurements on the local heliospheric magnetic field and the heliospheric
current sheet's tilt angle, to model the time-dependence and amplitude of cosmic-ray solar modulation. We find evidence 
for the solar modulation's charge- and rigidity-dependence during the era of solar 
cycle 24. Our analytic prescription to model solar modulation can explain well the large-scale time-evolution of positively 
charged cosmic-ray fluxes in the range of rigidities from 1 to 10 GV. We also find that cosmic-ray electron fluxes measured 
during the first years of cycle 24 are less trivial to explain, due to the complex and rapidly evolving structure of the 
Heliosphere's magnetic field that they experienced as they propagated inwards. 
\end{abstract}

\maketitle

\section{Introduction}
\label{sec:introduction}
The Sun produces a time-varying stream of charged particles known as the solar wind that extends 
out to at least 100 astronomical units (AU). This region is called the Heliosphere. 
The solar wind and its embedded magnetic field known as Heliospheric Magnetic Field (HMF) can have 
a strong effect on cosmic rays entering from the interstellar medium (ISM), the space outside the 
Heliosphere and between stellar systems in our Galaxy. Cosmic rays propagating through the Heliosphere 
get deflected and lose energy by interacting with the HMF. As the HMF changes with time its effect on the 
observed cosmic-ray spectra at Earth has an imprinted time-variation \cite{2011JGRA..116.2104U,  
Caballero_Lopez_2012, Corti:2018ycg,  Corti:2019jlt, Burger_2022}. This effect is known as solar modulation 
of cosmic rays. 
With current cosmic-ray observations \cite{Aguilar:2013qda, Adriani:2016uhu, Abe:2017yrg} the statistical errors associated
with the detected fluxes are now much smaller than the systematic uncertainties associated with cosmic-ray propagation, 
including solar modulation. Given this high precision era for cosmic rays, it is important to properly understand how 
the Sun influences these spectra in order to make reliable inferences on how cosmic rays are produced and 
propagate throughout the Milky Way \cite{Kachelriess:2011qv, Cholis:2013lwa, Hooper:2014ysa, Mertsch:2014poa, 
Strong:2015zva, Kohri:2015mga, Malkov:2016kbe, Cuoco:2016eej, Cui:2016ppb, Cholis:2017qlb, Cholis:2018izy, 
Lowell:2018xff, Kuhlen:2019hqb, PhysRevD.102.103007, Wang:2019xtu, Cholis:2019ejx, Cuoco:2019kuu, 
Heisig:2020nse, Engelbrecht_2021, Potgieter_2021, Cholis:2021kqk}. 

In this work, we follow a data-driven approach where we test the analytic model of \cite{Cholis:2015gna} to 
cosmic-ray measurements. This analytic model includes to first order the effects of cosmic-ray diffusion experienced 
through different regions of the Heliosphere and the presence of drift effects (see also \cite{Gieseler:2017xry,  
Aslam:2018kpi, Corti:2018ycg, Kuhlen:2019hqb, Vittino:2019yme}). 
To include the time-evolution of the HMF and its impact on solar modulation, we use ongoing observations 
from the Advanced Composition Explorer's (\textit{ACE})  \cite{1998SSRv...86..613S, ACESite} magnetometer 
and the Solar Wind Electron Proton Alpha Monitor (SWEPAM) \cite{SWEPAMsite}. We also use information 
on the value of the tilt angle of the heliospheric current sheet (HCS) -associated to the time-varying morphology 
of the HMF- from publicly available model-parameters provided by the Wilcox Solar Observatory  (WSO) \cite{WSOSite}.
We have found that the tilt angle of the heliospheric current sheet (HCS) and the magnitude of the HMF measured 
at the position of Earth have a strong correlation to the cosmic-ray solar modulation and are well observed 
using \textit{ACE} data and the WSO. However, we do not find a strong correlation to the solar 
wind's bulk speed essentially considering it fixed in time, while it still has a radial dependence 
\cite{2006SSRv..127..117H}.

As particles enter the Heliosphere they are deflected through the magnetic field.  For a given solar magnetic polarity $A$ of 
the HMF, depending on its charge $q$ a cosmic ray particle is more likely to reach the Earth by propagating 
through different regions of the Heliosphere. When the combination of $qA <$ 0, particles 
propagate mostly through the HCS, while when $qA >$ 0, particles propagate mostly through the 
solar magnetic poles \cite{1981ApJ...243.1115J, 1983ApJ...265..573K, 2004JGRA..109.1101C, Potgieter:2013pdj, 
POTGIETER20141415, POTGIETER2017848}. Moreover, particles that propagate through the HCS travel 
slower compared to particles traveling through the magnetic poles with the same magnitude of rigidity 
$|R| = p/|q|$ (where $p$ is the particle's momentum and $q$ its charge). 
This results in larger energy losses for the former particles 
on average and thus a more significant change in their observed intensity and energy. In our work, we model the 
averaged effect of the different paths that cosmic rays of opposite charge follow as they propagate to the Earth. 
We note that this assumption effectively breaks the spherical symmetry that conventional force field
models assume for the solar modulation \cite{1968ApJ...154.1011G}. Extensive numerical works have also explored this 
break away from 1-dimensional propagation \cite{2004JGRA..109.1101C, 2012Ap&SS.339..223S, Qin2017, 2018ApJ...858...61B, 
2019arXiv190307501J, Caballero_Lopez_2019, 2019ApJ...878...59B}. However, as we will describe in detail we retain the force field 
notation as it gives a simple qualitative description of the impact the HMF has on the cosmic-ray spectra measured
at Earth versus those in the local ISM.

This paper is organized as follows; in Section~\ref{sec:Data} we present the observations we use to model the time-varying HMF
properties and also the cosmic-ray spectra that we rely on. In Section~\ref{sec:Method}, we present the analytic
model for the solar modulation that we test to the cosmic-ray measurements. In Section~\ref{sec:Results}, we present
our results finding that indeed there is a clear indication in the cosmic-ray data of a charge dependence on the
solar modulation. Our charge-, time- and rigidity-dependent model can explain well the larger-scale time-evolution 
of positively charged cosmic-rays (hydrogen and positrons) in the range of rigidities from 1 to 10 GV. We can also 
explain some of the observations of electrons in the same rigidity range and time, especially once we find that cosmic-rays experience a stable Heliospheric
polarity  as they propagate inwards. Finally, in Section~\ref{sec:Conclusions} we present
our conclusions and discuss the remaining open questions and further directions on the modeling of the 
time-varying, change- and rigidity-dependent effects of solar modulation.

\section{Data}
\label{sec:Data}
\begin{figure}
\begin{center}
\hspace{-0.6cm}
\includegraphics[width=3.55in,angle=0]{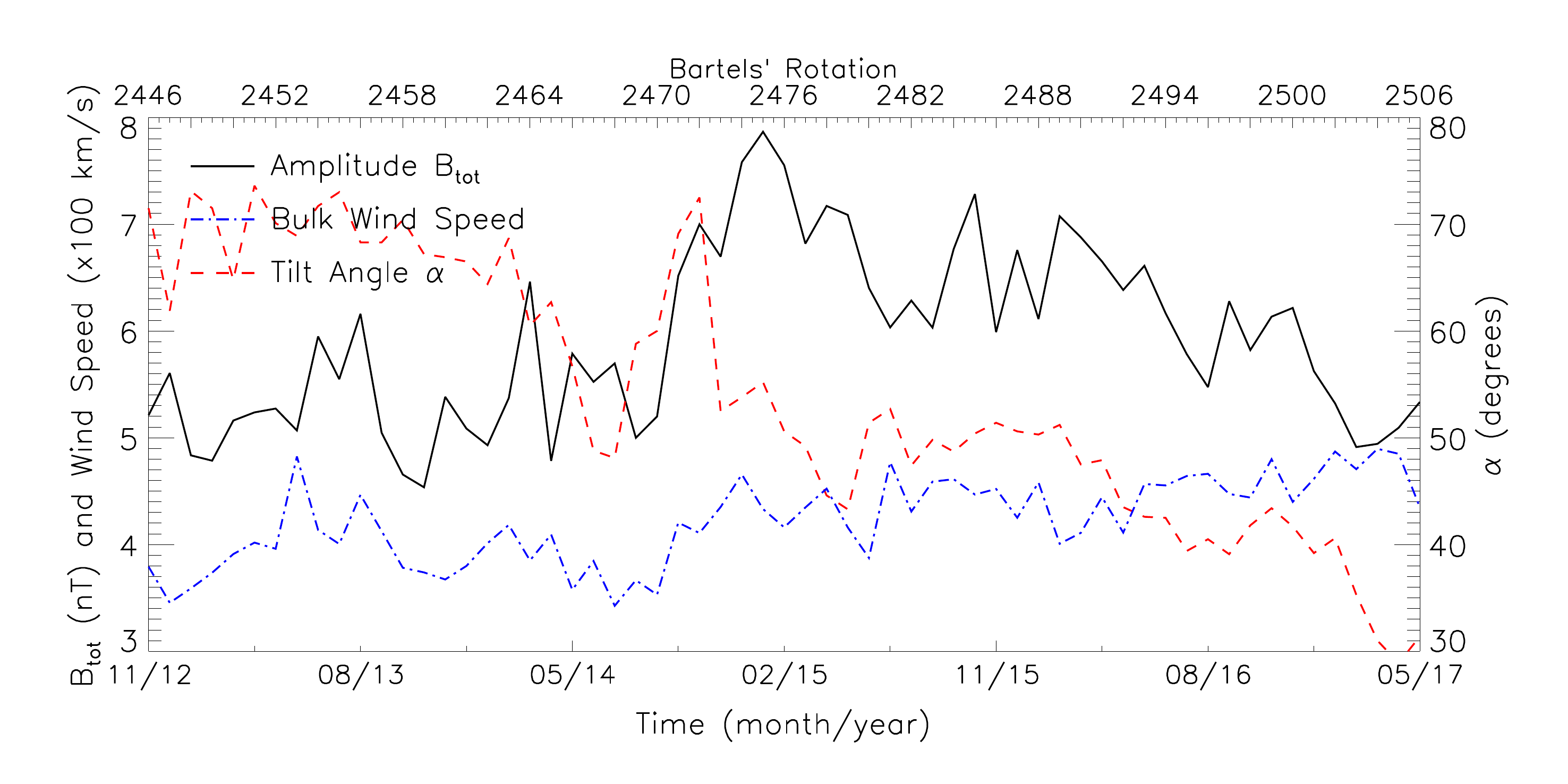}
\end{center}
\vspace{-0.6cm}
\caption{The amplitude of the solar magnetic field $|B_{tot}|$, measured at 1 AU (solid black) with axis units 
and range information on the left $y$-axis. Also plotted are the solar wind bulk speed at 1 AU (blue dashed-doted 
line) with its units and range also on the left $y$-axis and the tilt angle of the current sheet (red dashed line 
and right $y$-axis).}
\label{fig:Tilt_BField}
\end{figure}

This paper covers observations made during solar cycle 24, and in particular we focus on the era from 
Bartels' rotation (BR) cycle 2456 to Bartels' rotation cycle 2506 (roughly August 2013 to April 2017). 
During this era, the properties of the HMF have been extensively studied and measured. It represents also a 
time interval during which the polarity of the HMF is well defined to be positive ($A>0$). 
In Fig.~\ref{fig:Tilt_BField}, the amplitude of the HMF $|B_{tot}|$, measured at the Earth's location (at 1 AU), 
the tilt angle $\alpha$, and the solar wind's bulk speed measured also at 1 AU are plotted as they change 
over time. While the strength of the magnetic field and the bulk speed are directly measured by the 
ACE instruments, the plotted tilt angle gives the maximum extent in latitude reached by the 
computed HCS, i.e. describes the general structure of the HMF. This is a model-dependent result 
\footnote{We follow the ``classic'' model's values for the tilt angle from \cite{WSOSite}, as it has been suggested to give a more
accurate estimate \cite{FERREIRA2003657}.}, relying on observations of the solar magnetic field.
The time-varying tilt angle information is publicly available from the WSO 
\cite{WSOSite, 1992sws..coll..191H, 2004ApJ...603..744F}.
To first order the bulk wind speed time-variation is very mild and from here on its time-dependence 
is ignored. We concentrate on the HMF's $|B_{tot}|$ and $\alpha$ quantities, in agreement 
with earlier works \cite{Cholis:2015gna, Cholis:2020tpi}.

We use cosmic-ray observations made by the Alpha Magnetic Spectrometer (\textit{AMS-02}) to fit the 
model parameters of Ref.~\cite{Cholis:2015gna}. In particular, we use the cosmic-ray hydrogen (proton plus 
deuteron) spectra taken from 2013 to 2017, between BR 2456 and BR 2506 \cite{2018PhRvL.121e1101A} and 
the equivalent period for electron and positron cosmic-ray spectra \cite{2018PhRvL.121e1102A}. These observations 
by \textit{AMS-02} allow us to model changes between Bartels' rotation cycles in the respective cosmic-ray fluxes i.e. 
changes that appear over 27-day long time scales. Moreover,  to study the rigidity-dependence of solar modulation, 
we use observations over five separate rigidity ranges. Those span the ranges of 1.16-1.33, 1.92-2.15, 3.29-3.64, 
5.37-5.90, and 9.26-10.1 GV. We do not expand our analysis to higher rigidities as the respective solar modulation 
changes between BRs are too small. Once studying the observations 
of that era we noticed that the cosmic-ray fluxes fluctuate roughly around the same average over the first half of 
2013-2017 era, while after that all fluxes increase nearly monotonically. For that reason, we break our analysis 
in two sub-eras, the first one being from BR 2456 to BR 2481 and the second from BR 2482 to BR 2506. Since 
we care about the evolution of the fluxes, we follow Ref.~\cite{Cholis:2020tpi}, where it was shown that in order 
to reduce the impact of ISM assumptions, it is best to study the evolution of the fluxes around their respective 
averaged flux. This still allows us to focus on the relative 
changes of the hydrogen, electron and positron fluxes over time.
We also note that at lower rigidities the time evolution of the cosmic-ray hydrogen flux -and thus also its ratio to 
the averaged observed flux- is significantly more prominent. At the 1.16-1.33 GV bin the time variation is at the 
$-50\% \;+80\%$ level for the entire 2013-2017 era, while at the 9.26-10.1 GV bin it is at the $\pm 5\%$ level. 

\section{Methodology}
\label{sec:Method}
The analytic treatment of solar modulation that we employ follows \cite{Cholis:2015gna}, where the force-field 
approximation was expanded to include a time-, charge- and rigidity-dependence on the solar modulation. 
The effect of solar modulation  like in \cite{1968ApJ...154.1011G}, is that the kinetic energy $E_{kin}$ of each 
particle is reduced on average by $|Z|e\Phi$, where $\Phi$ is the modulation potential, generally on the order 
of 0.1 - 1.0 GV, and $|Z|e$ is the absolute charge of the cosmic ray. The resulting effect of the modulation 
potential on the cosmic-ray differential spectrum can be written as:
\begin{eqnarray}
  \frac{dN^{\oplus}}{dE_{kin}}(E_{kin}) &=& \frac{(E_{kin} + m)^2 - m^2}{(E_{kin} + m + |Z|e\Phi)^2 - m^2}   \nonumber \\
  &\times& \frac{dN^{ISM}}{dE_{kin}}(E_{kin} + |Z|e\Phi)
  \label{eq:Ekin_at_Earth}
\end{eqnarray}
where $E_{kin}$ is the observed kinetic energy and the subscripts ``$ISM$" and ``$\oplus$" denote the respective 
values in the local interstellar medium and at the location of Earth. In the standard force field approach, the value of 
$\Phi$ is fitted to the cosmic-ray observations without a prediction on what its value should be at any given time nor 
accounting for the fact that particles of opposite charge propagate through different regions of the Heliosphere. 
Moreover, the value of $\Phi$ is assumed to be the same for all cosmic rays irrespective of the energy they carry as they 
enter the HMF. In this work we instead follow \cite{Cholis:2015gna}, where the solar modulation potential depends on 
three well-studied quantities: the polarity of the solar magnetic field, the magnitude of the HMF at 1 AU, and the 
tilt angle of the HCS. The analytic expression for the solar modulation potential is as follows:
\begin{eqnarray}
  \Phi(R, q, t) &=& \phi_{0}\left(\frac{|B_{tot}(t)|}{4 nT}\right)  \nonumber \\ 
  &+& \phi_{1}H(-qA(t)) \left(\frac{|B_{tot}(t)|}{4 nT}\right) \nonumber \\  
  &\times& \left(\frac{1+(\frac{R}{R_{0}})^2}{\beta(\frac{R}{R_{0}})^3}\right)\left(\frac{\alpha(t)}{\frac{\pi}{2}}\right)^4.
\label{eq:ModPot}
\end{eqnarray}
$A$ is the polarity of the HMF and $|B_{tot}|$ its strength as measured at Earth. $\alpha$ is the tilt angle of the HCS. 
These quantities are considered time-dependent inputs, and therefore independent of cosmic-ray observables. 
$R$, $\beta$, and $q$ are the rigidity, velocity/$c$, and 
charge of the cosmic ray, respectively. $H$ is the Heaviside step function separating the treatment of opposite charges 
cosmic rays based on the product of $q$ and $A$ as in Figure~1 of \cite{Cholis:2015gna}. $R_{0}$ is a reference 
rigidity at which point drift effects are important along the HCS and $\phi_{0}$ and $\phi_{1}$ are normalization 
factors fitted to the data that represent the amplitude of the combined effect of diffusion and drift. To test our solar 
modulation model to cosmic-ray measurements conducted at different times we use \textit{AMS-02} data from Refs.~\cite{2018PhRvL.121e1101A,
2018PhRvL.121e1102A}, as described in Section~\ref{sec:Data}.
We constrain the parameters of $\phi_{0}$, $\phi_{1}$ and $R_{0}$ by fitting them to the available cosmic-ray data 
for solar cycle 24.

\begin{figure*}
\begin{center}
\begin{tabular}{c c}
\vspace{-0.08in}
\hspace{-0.15in}
\includegraphics[width=3.72in,angle=0]{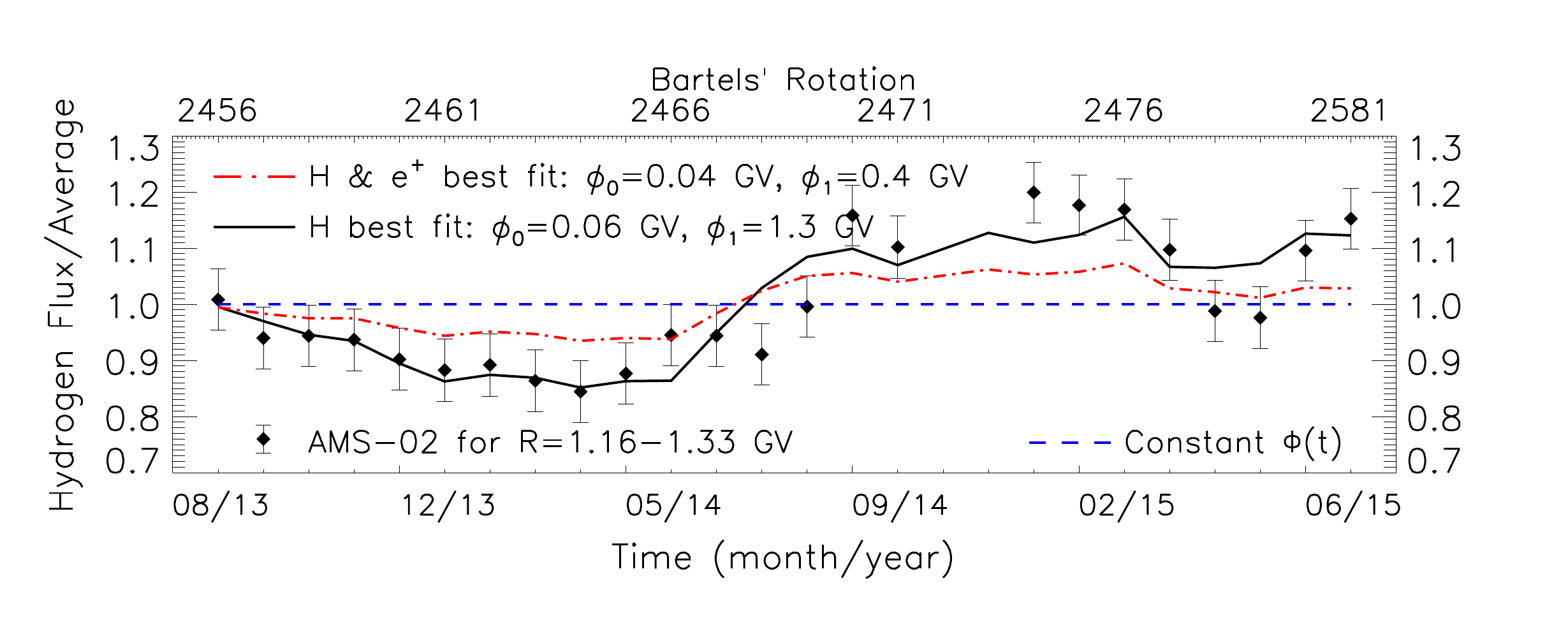}
\hspace{-0.15in}
\includegraphics[width=3.72in,angle=0]{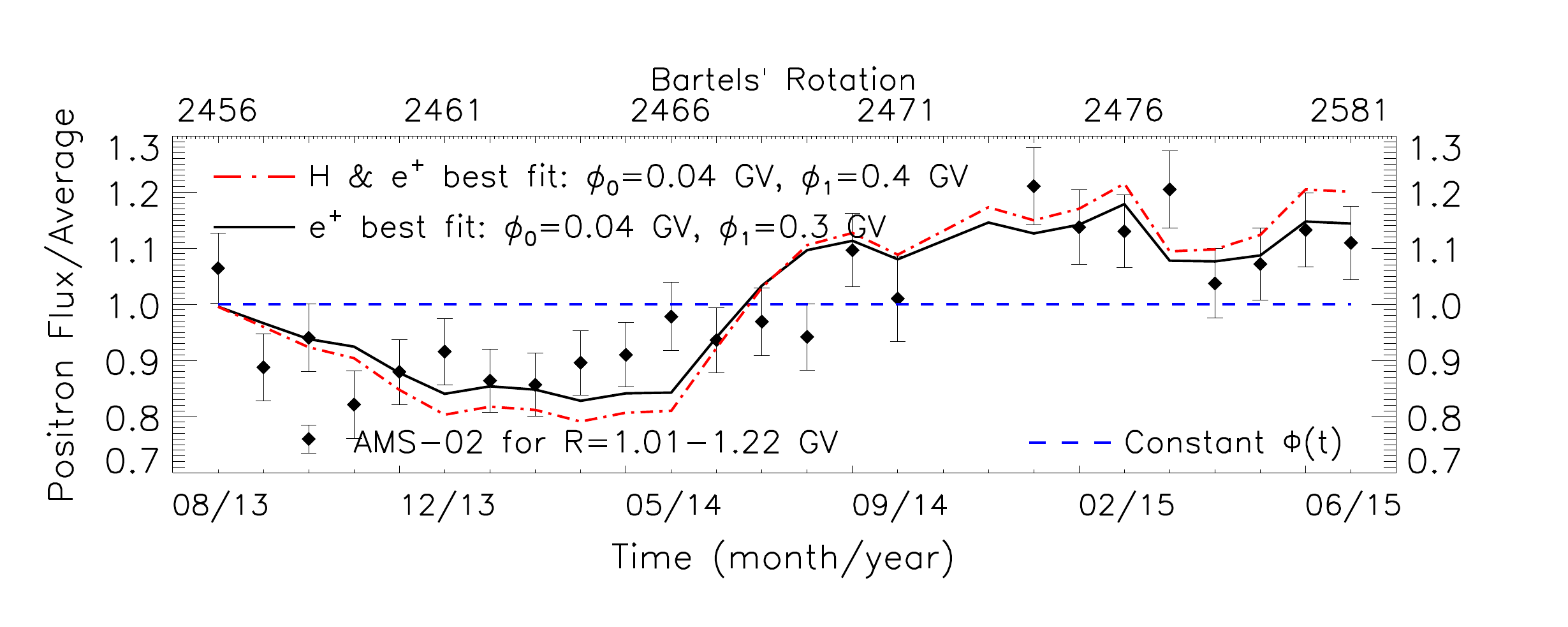}\\
\vspace{-0.08in}
\hspace{-0.15in}
\includegraphics[width=3.72in,angle=0]{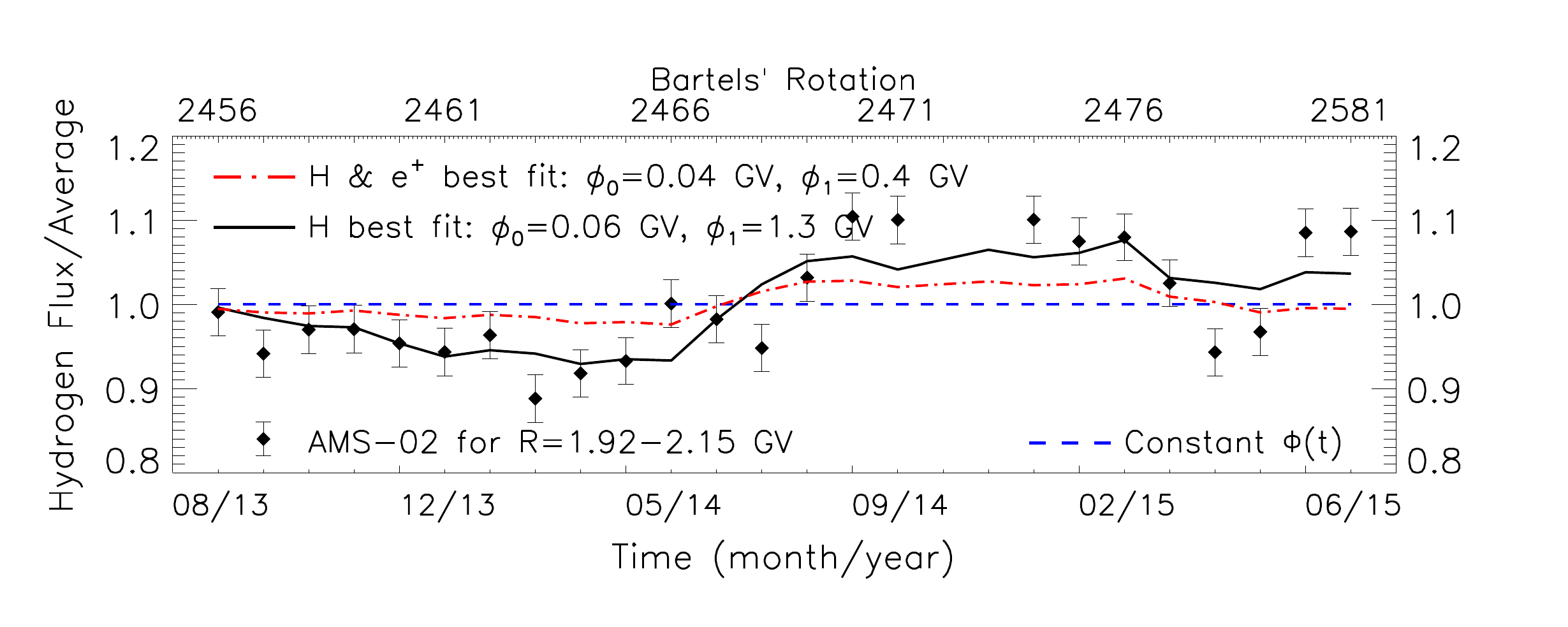}
\hspace{-0.15in}
\includegraphics[width=3.72in,angle=0]{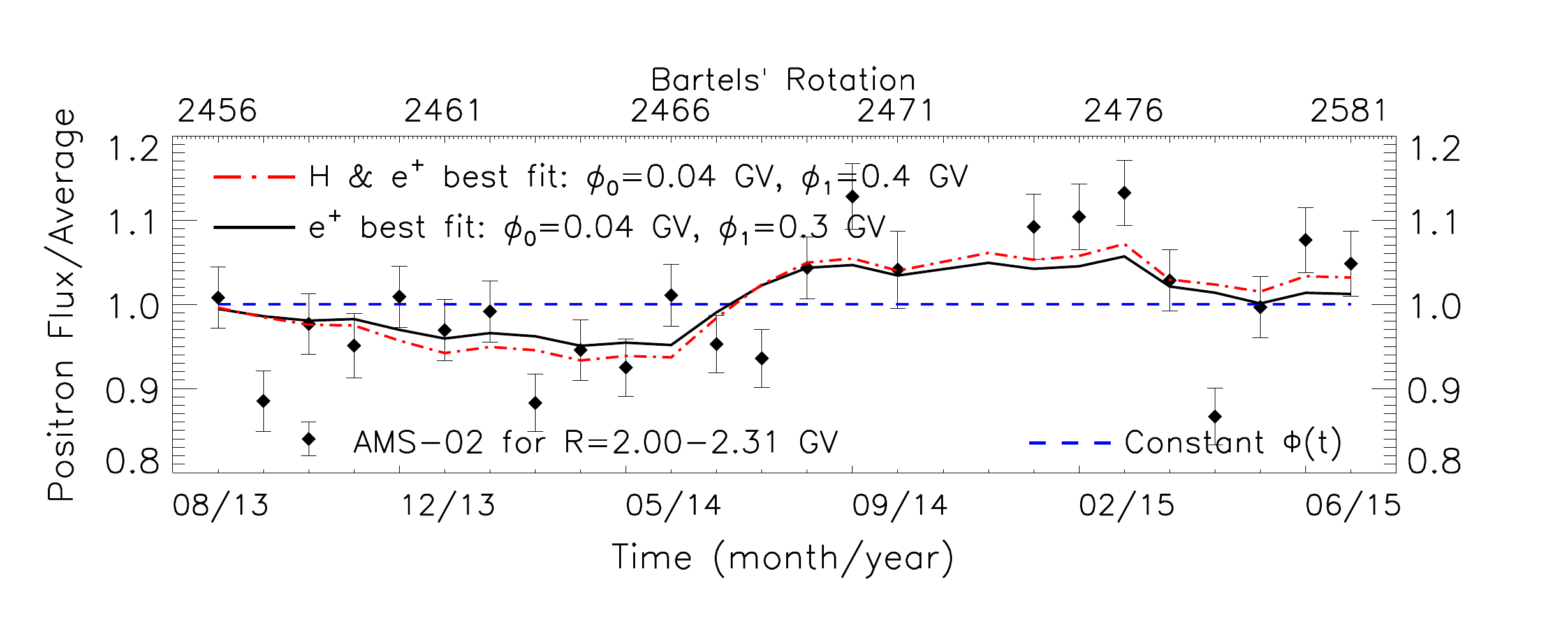}\\
\vspace{-0.08in}
\hspace{-0.15in}
\includegraphics[width=3.72in,angle=0]{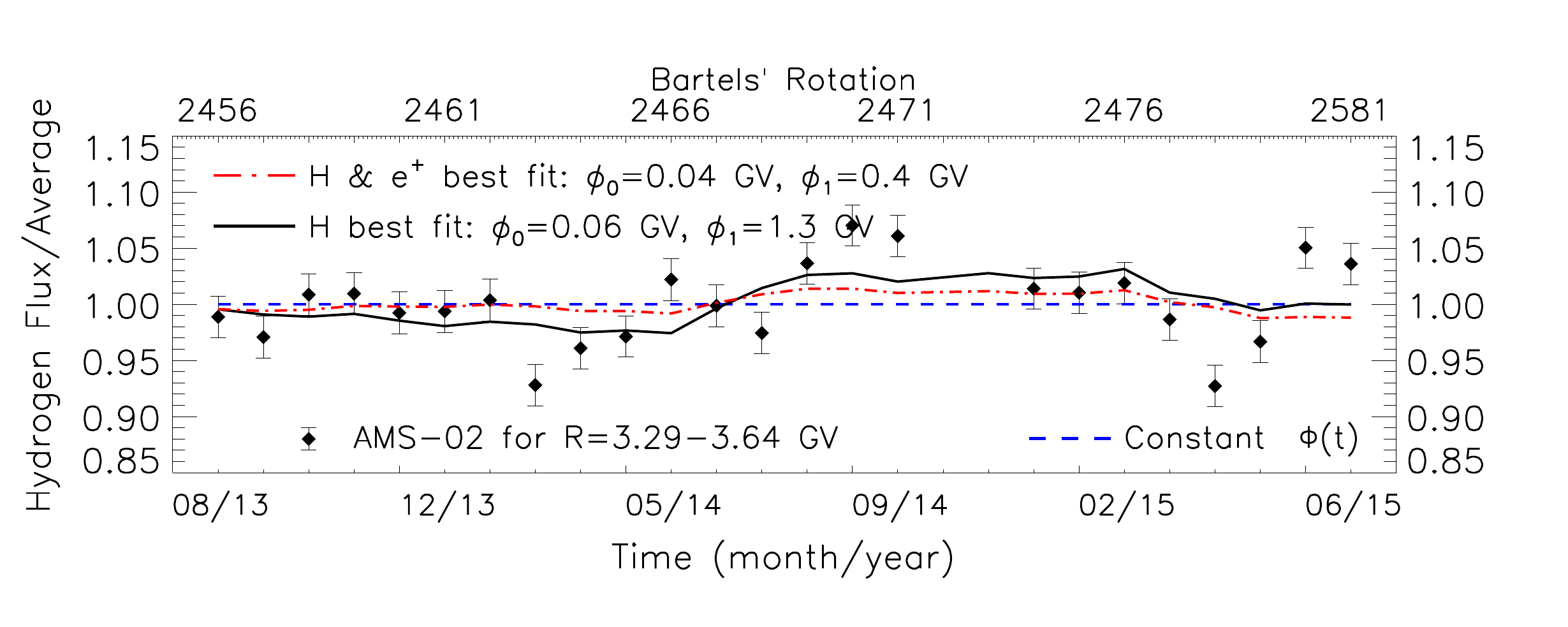}
\hspace{-0.15in}
\includegraphics[width=3.72in,angle=0]{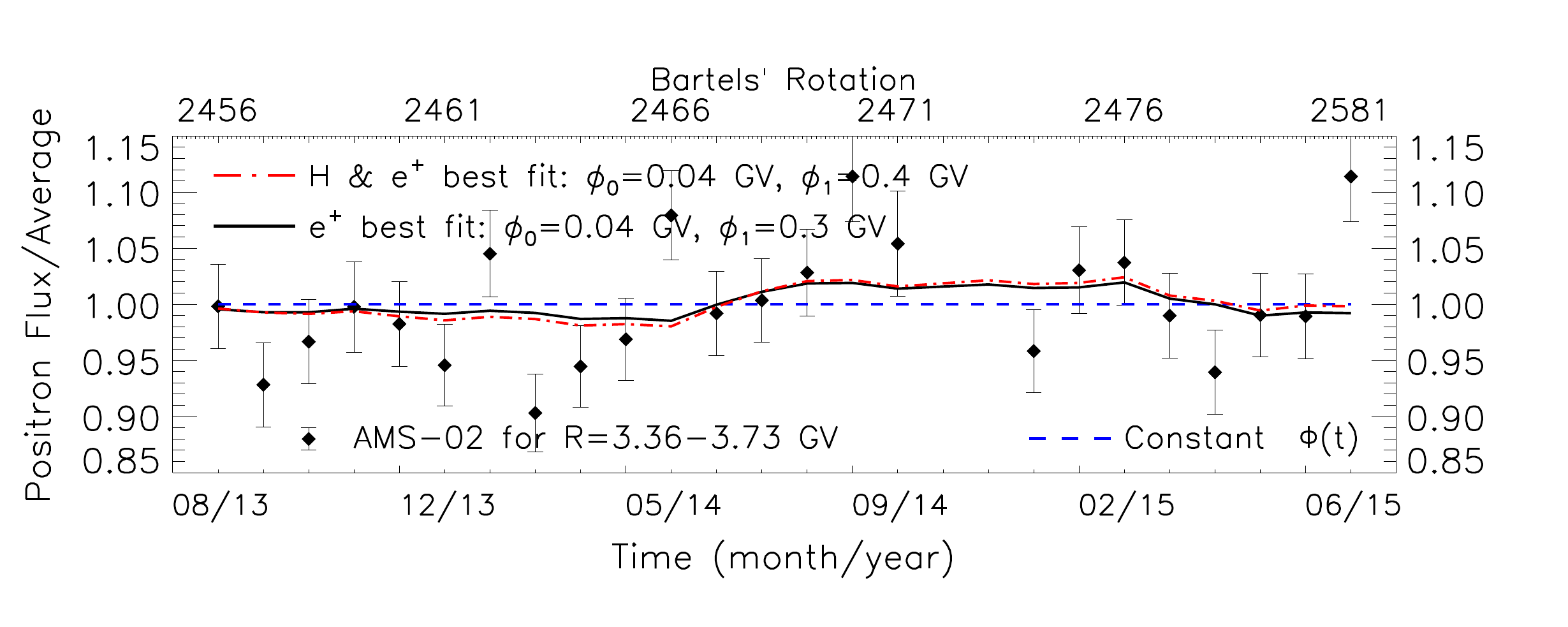}\\
\vspace{-0.08in}
\hspace{-0.15in}
\includegraphics[width=3.72in,angle=0]{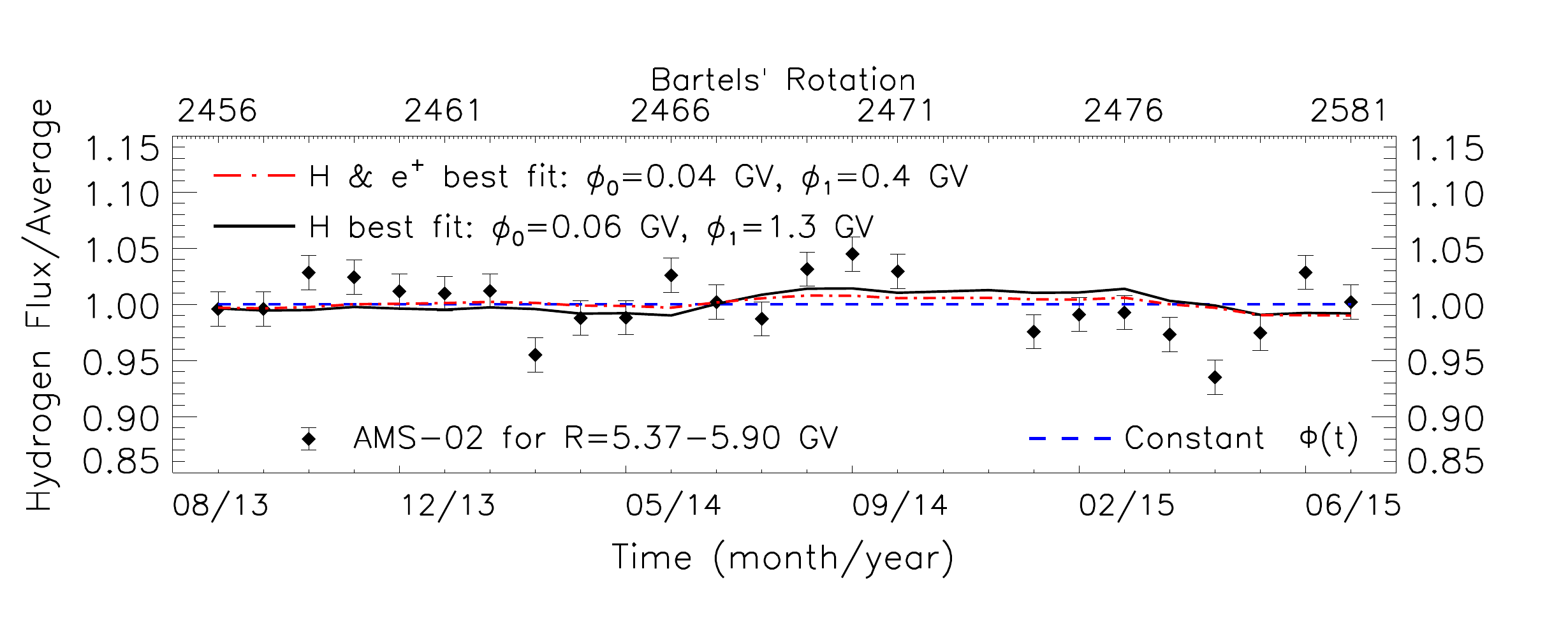}
\hspace{-0.15in}
\includegraphics[width=3.72in,angle=0]{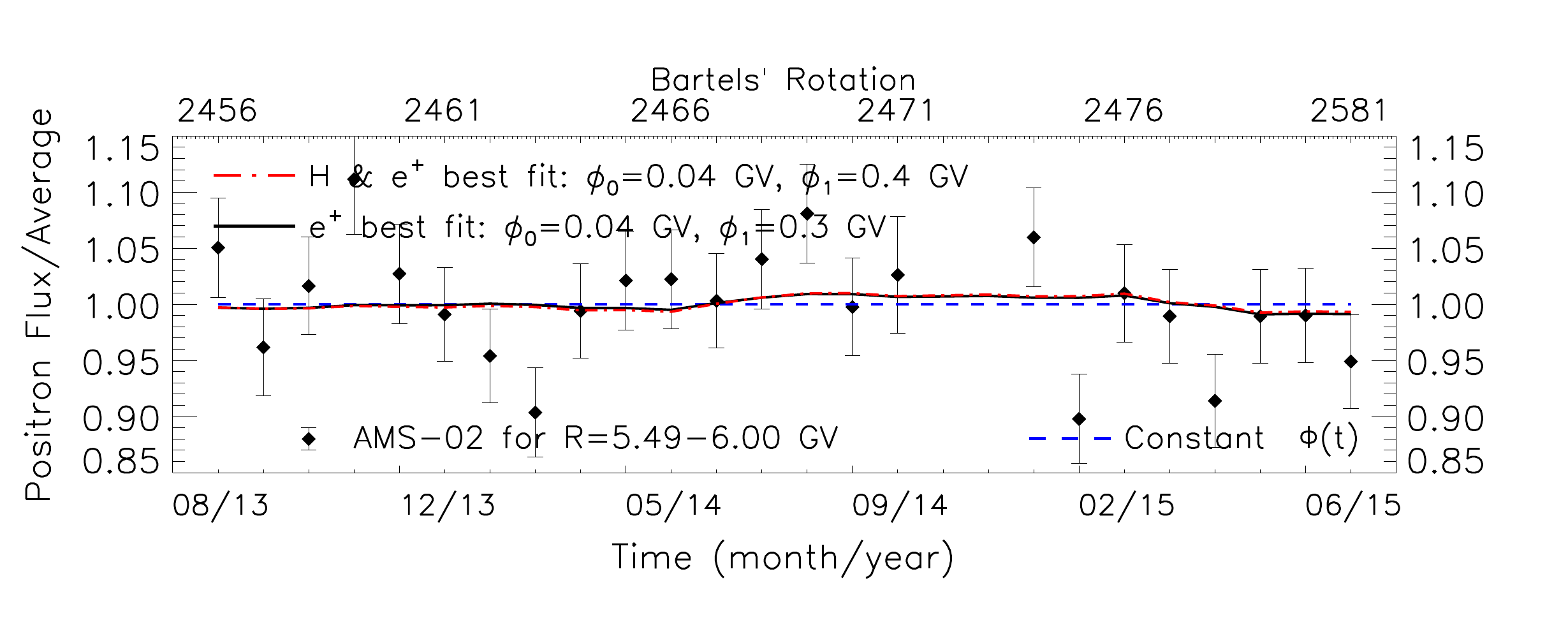}\\
\vspace{-0.08in}
\hspace{-0.15in}
\includegraphics[width=3.72in,angle=0]{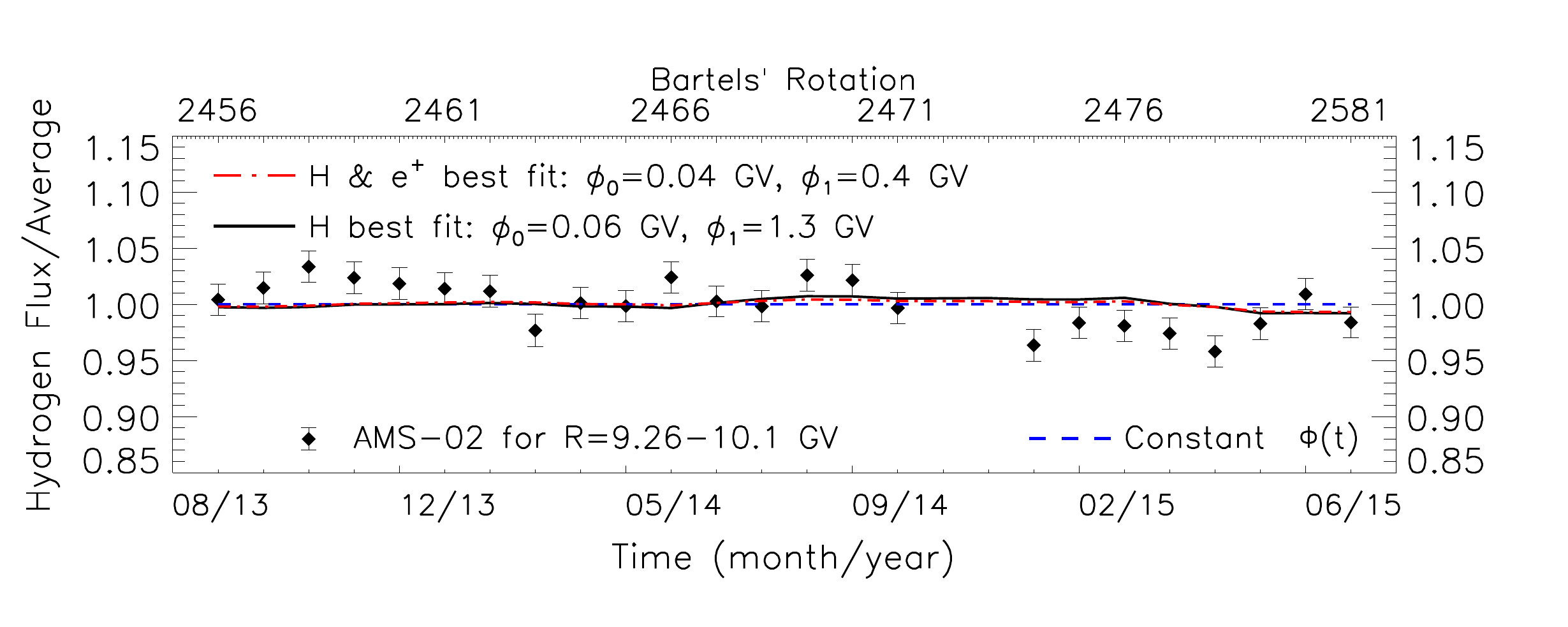}
\hspace{-0.15in}
\includegraphics[width=3.72in,angle=0]{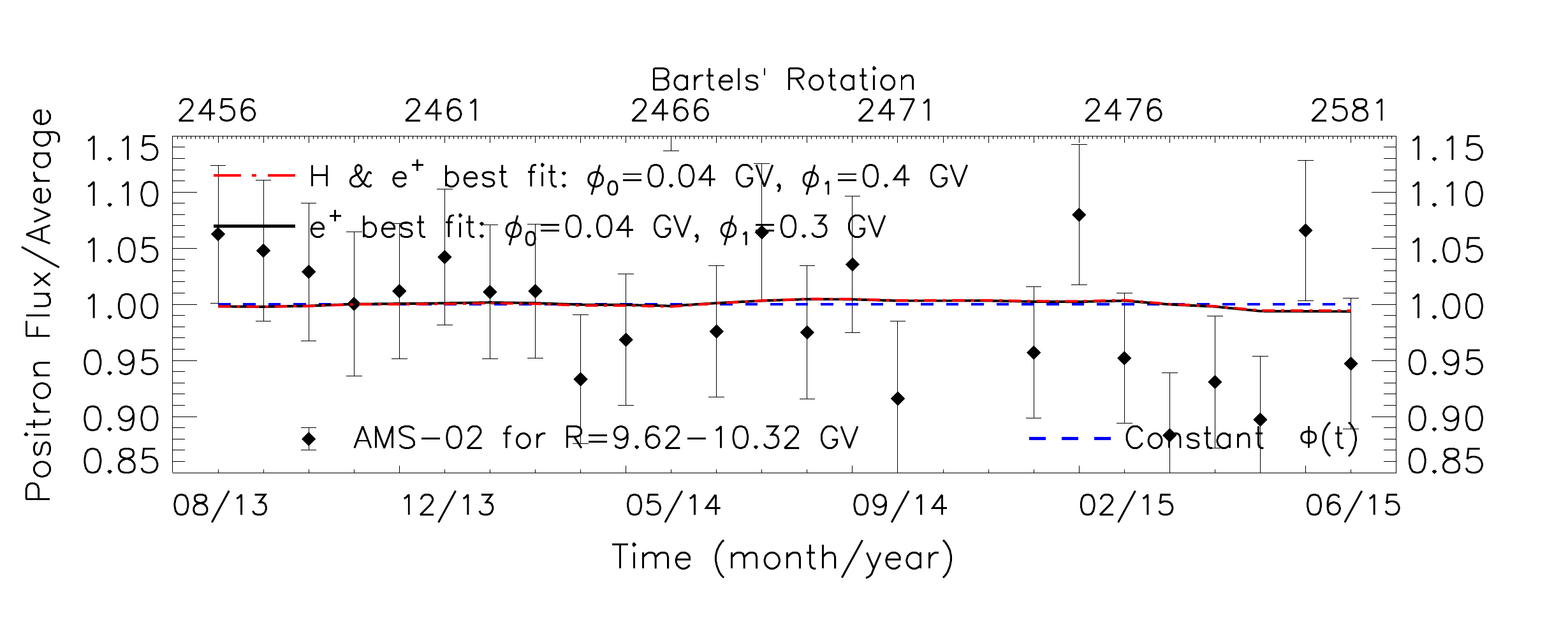}\\
\end{tabular}
\end{center}
\vspace{-0.15in}
\caption{\textit{Left column}: the time evolution of the ratio of the cosmic-ray hydrogen flux to the averaged 
hydrogen cosmic-ray flux within a period of the 24 observed Bartel's cycles (BR 2456-2481). We show the 
following five rigidity bins, from top to bottom: 1.16-1.33, 1.92-2.15, 3.29-3.64, 5.37-5.90, 9.26-10.1 GV. The 
error-bars come from the \textit{AMS-02} observations (see text for details), while the black line gives the best 
fit choice for the hydrogen (H) ratio for $\phi_{0} = 0.06$ GV, $\phi_{1} = 1.3$ GV and assuming $R_{0} = 0.5$ GV.
\textit{Right column}: the time evolution of the ratio of the cosmic-ray positron flux to the averaged positron 
cosmic-ray flux within the same period of the 24 observed Bartel's cycles. The rigidity bins used are not identical
to the hydrogen ones as the rigidity binning performed by \textit{AMS-02} is not the same for the two cosmic-ray species. 
We used the closest possible positron rigidity bins of 1.01-1.22, 2.00-2.31, 3.36-3.73, 5.49-6.00, 9.62-10.32 GV. The black line 
gives the best fit choice for the positron ($e^{+}$) ratio for $\phi_{0} = 0.04$ GV, $\phi_{1} = 0.3$ GV and assuming 
$R_{0} = 1.0$ GV. The red dashed-dotted lines on both the left and right columns give the evolution of the respective 
hydrogen and positron flux ratios as predicted for values of $\phi_{0} = 0.04$ GV and $\phi_{1} = 0.4$ GV.
These values give the best fit to the combination of the positively charged cosmic-ray particles used for that
era (see text for further details).}
\label{fig:PositiveCharges_BR2456-2481}
\end{figure*}

We perform a $\chi^{2}$ test  to get the best fit values of the parameters, $\phi_{0}$, $\phi_{1}$ and $R_{0}$.
This parameter space is probed through a discrete grid that was found to give smoothly changing $\chi^{2}$ values. 
For each combination of $\phi_{0}$, $\phi_{1}$ and  $R_{0}$ values, the solar modulation function is calculated for 
every Bartels' cycle, once for protons and once for deuterons. The resultant fluxes of protons and deuterons are 
added to get a value for the hydrogen flux at each Bartels' cycle. The hydrogen fluxes over the 24 Bartels' cycles
\footnote{\textit{AMS-02} does not provide measurements for BR 2472 and BR 2473, so these two cycles are omitted 
from the fit even though our model does provide a prediction as shown in our figures.} 
of BR 2456 to 2581 are averaged for each of the five rigidity bins of 1.16-1.33, 1.92-2.15, 3.29-3.64, 5.37-5.90, and 
9.26-10.1 GV. Then, the value for the hydrogen flux at each Bartels' cycles is divided by the average hydrogen flux, 
resulting in a list of 24 hydrogen flux ratios, that act as the ``expected'' values of hydrogen flux ratio for the given 
parameter values. Each of those ratios has an uncertainty that is directly proportional to the reported \textit{AMS-02}
flux uncertainty (statistical and systematic added in quadrature) for the given Bartels' cycle. Those ratios and respective 
uncertainties are represented by the ``\textit{AMS-02}'' data with error-bars in Fig.~\ref{fig:PositiveCharges_BR2456-2481} 
in the left column for the hydrogen flux ratio (``Hydrogen Flux/Average''). We repeat the same procedure for positrons 
(shown in Fig.~\ref{fig:PositiveCharges_BR2456-2481} right column), as well as electrons for BR 2456-2481 and again 
for the hydrogen, positron and electron observations of the era between BR 2482 and BR 2506 as we describe in 
in Section~\ref{sec:Results}.

By taking the ratio of the cosmic-ray hydrogen or positron or electron fluxes at different Bartels' rotations over their
respective average we find that some of the underlying systematic modeling uncertainties of the ISM fluxes cancel out, 
allowing us for a more direct test of solar modulation (see also \cite{Cholis:2020tpi}). Moreover, the local cosmic-ray ISM 
fluxes in that range of rigidities i.e. 1-10 GV are stable on timescales of order $O(10)$ Myr allowing only for the solar 
modulation to cause any changes. In fact, even at rigidities as high as TVs, local sources such as pulsars or recent 
supernova remnants can have an effect on the spectra on a timescale of only $O(10^4)$ years (see e.g. 
\cite{Profumo:2008ms, Cholis:2013lwa, Mertsch:2014poa, Cholis:2017qlb, Cholis:2018izy, Mertsch:2018bqd, Cholis:2021kqk}). 

The standard $\chi^{2}$ test is performed, comparing the list of ``expected'' hydrogen, or positron, or electron flux 
ratios to the observed cosmic-ray spectra from \textit{AMS-02}, summing up the value for every Bartels' cycle to get 
the overall $\chi^{2}$ value. 

For the predicted local ISM fluxes we used the predictions of model ``C'' from \cite{Cholis:2021rpp} for the protons 
and deuterons. These ISM assumptions were found to give cosmic-ray fluxes in agreement with the hydrogen, 
helium, and carbon fluxes as well as boron-to-carbon, carbon-to-oxygen and beryllium-to-carbon flux ratios measured by 
\textit{AMS-02} at the entirety of the energies not affected by solar modulation. These cosmic-ray spectra have been 
evaluated by running the publicly available \texttt{GALPROP} code \cite{galprop, GALPROPSite, NEWGALPROP}. 
In addition, the local ISM positron and electron fluxes come from \cite{Cholis:2021kqk} (``Model I'') that relies on 
the ISM assumptions of  model ``C'' from \cite{Cholis:2021rpp} for the calculation of primary and secondary 
cosmic-ray electron and positron fluxes and, through the presence of local pulsars, further explains the 
cosmic-ray positron fraction ($e^{+}/(e^{+} + e^{-})$) spectrum, cosmic-ray positron, cosmic-ray electron plus positron flux 
observations from \textit{AMS-02} and from the Dark Matter Particle Explorer (\textit{DAMPE}) and the CALorimetric 
Electron Telescope (\textit{CALET}) \cite{AMS:2019rhg, AMS:2019iwo, AMS:2021nhj, DAMPE:2017fbg, Adriani:2018ktz}.

\section{Results}
\label{sec:Results}

Positively charged particles during cycle 24 mostly probe the parameter $\phi_{0}$, as during that time $A>0$ 
and they propagate to the Earth through the heliospheric poles. Conversely, negatively charged particles probe  
the $\phi_{1}$ and $R_{0}$ parameters as well. That can be seen in Eq.~\ref{eq:ModPot} through its switching off or on 
of the second term on the total modulation potential $\Phi(R, q, t)$. For that reason, we test the time evolution of 
the cosmic-ray hydrogen, positron and electron fluxes. Yet, that statement is accurate only for the case where 
the observed cosmic-rays measured at Earth propagated through the Heliosphere while the polarity of the HMF 
was well established. This is important as from simulations (see e.g. Ref. \cite{2012Ap&SS.339..223S}) we 
understand that cosmic-ray particles may take approximately a year for them to propagate through the $\simeq 100$ 
AU of distance to reach the \textit{AMS-02} detectors. Moreover, given the bulk speed of the outward moving 
solar wind, it takes in addition half a year to a year for the HMF new polarity to be established all the way to the edge 
of the Heliosphere. Thus, even if the HMF polarity $A$ flip is instantaneous at the Sun, cosmic-ray particles reaching 
the Earth two years after that moment may have experienced a non-well defined HMF polarity during their inward 
propagation. There is a significant time-delay between HMF changes observed at the surface of the Sun or at 1 AU 
and their effect on cosmic-ray fluxes through solar modulation. In Ref.~\cite{Cholis:2020tpi}, that time delay was 
estimated to be $\simeq 20$ months even when studying observations made at the end of solar cycle 23 before the 
change of HMF from $A<0$ to $A>0$. In our case that time-delay may be even larger especially since the polarity 
flip was not instantaneous.  Based on observations reported by the WSO, the most likely moment of the polarity 
flip was May of 2013. However, the Sun's 30-day window averaged polar field measurements between October 
2012 and March 2014 as reported in \cite{WSO_polarity} suggest that the polarity flip of the Sun's field 
was not instantaneous. 

Originally we started to study data from BR 2453 onwards (May 2013) and averaged the entire era of BR 2453 
to BR 2506 when the \textit{AMS-02} observations of Refs.~\cite{2018PhRvL.121e1101A, 2018PhRvL.121e1102A} end 
\footnote{\textit{AMS-02} observations from the era of cycle 23 taken between BR 2426 to 2447 (May 2011 to 
December 2012) were studied in Ref.~\cite{Cholis:2020tpi} and we are going to draw comparisons to that work in 
our Section~\ref{sec:Conclusions}.}. 
We noticed two important points. One, it is very difficult to explain the entirely of that time with any simple model
dependent on the polarity, tilt angle, total magnetic field or bulk speed observations. Moreover, the first months 
are too close to the polarity flip to be a useful to understanding the overall propagation of cosmic rays through 
the Heliosphere. Simply put, those cosmic rays as they were approaching the Earth experienced a significant 
field change towards the end of their path. 

\begin{figure}
\begin{center}
\hspace{-0.6cm}
\includegraphics[width=3.55in,angle=0]{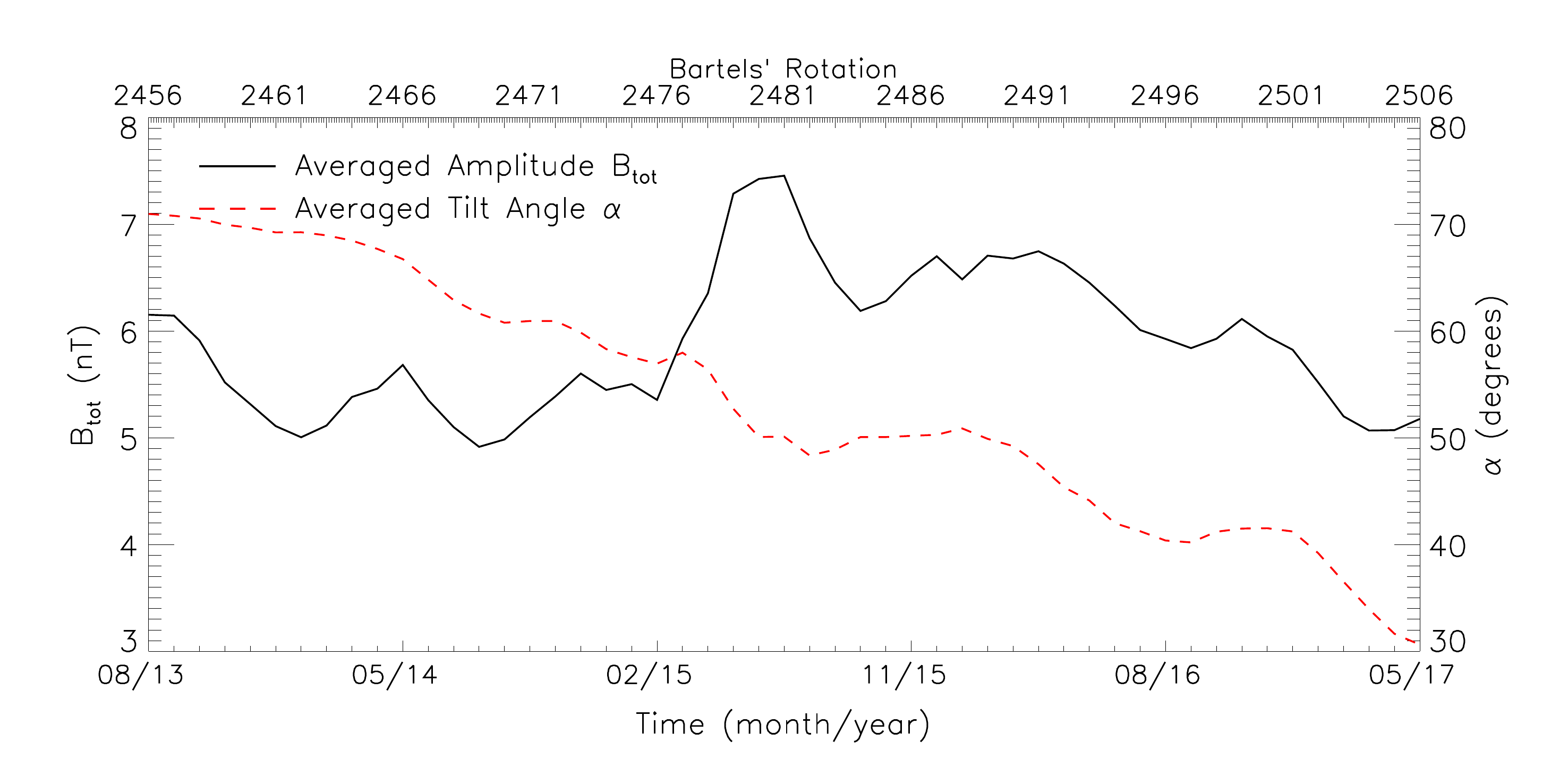}
\end{center}
\vspace{-0.6cm}
\caption{Time evolution of the \textit{averaged} HMF properties that cosmic rays at Earth have experienced.}
\label{fig:AveragedHMF}
\end{figure}

We chose to focus on the data from BR 2456 onwards.  In 
addition, we break the data into two sections of equal time intervals of BR 2456-2481 when the effect of a non-well 
defined polarity experienced by the inward propagated observed cosmic rays suggests the presence of both the 
$\phi_{0}$ and $\phi_{1}$ terms of Eq.~\ref{eq:ModPot} on positively and negatively charged particles. However, 
for BR 2482-2506, positively charged particles experience a monotonic turning-off the $\phi_{1}$-term which 
completely shuts down at BR 2490. This is in agreement with the long time delay between the polarity change 
on the Sun's surface and its effect on the observed cosmic rays. For the eras that we study, we provide in 
Fig.~\ref{fig:AveragedHMF}, the averaged HMF magnetic field amplitude (using its value at 1 AU as a probe for
time-evolution in the Heliosphere) and tilt angle that the cosmic rays experience. As can be seen small timescale 
effects of Fig.~\ref{fig:Tilt_BField} are washed out. 

For particles arriving at Earth up to BR 2481 we took the same averaging assumptions as in Ref.~\cite{Cholis:2020tpi}. 
That is the $B_{\rm tot}(t)$ is averaged over 4 BRs with a time delay of 16 BRs. i.e. for positively charged particles 
arriving at BR 2456 we evaluate $B_{\rm tot}(t)$ of Eq.~\ref{eq:ModPot}  using the \textit{ACE} observations of BR 
2437-2441. For the $\alpha(t)$ we use instead the averaged value of the last 20 BRs, without any further time-delay, 
i.e. in the specific example the 
WSO modeled predictions for $\alpha$ between BR 2437-2456. These choices where shown to provide the best fit to 
the \textit{AMS-02} observations of positively charged particles traveling through the HCS and are true even for the first 
Bartels' rotations with the new solar cycle. We did confirm that by testing alternative averaging choices and came to the 
same conclusions as in~\cite{Cholis:2020tpi}. For later Bartels' rotations, instead these same particles traveled through 
the Heliospheric poles; thus traveling faster to the Earth. Thus their averaging scheme is expected to be shorter 
(see e.g. \cite{2012Ap&SS.339..223S, Potgieter:2013pdj}). In Table~\ref{tab:Time_delay_fits} we give the assumptions 
that we tested for the positively charged particles for the period where we expect that cosmic rays propagated  
inwards through the poles. We used the era of BR-2482-2506. We report the difference in the $\chi^{2}$ fit of the five
rigidity bins for the hydrogen data that we use ($\Delta \chi^{2}$), between the best choice and alternative ones. The
best fit choice is achieved with an averaging scheme where both $B_{\rm tot}(t)$ and $\alpha(t)$ are averaged over 
4 BRs with zero time delay. That is, for a positively charged particle that arrived at BR 2500 the evaluated $B_{\rm tot}(t)$ 
and $\alpha(t)$ of Eq.~\ref{eq:ModPot}, used the \textit{ACE} and WSO averaged observations of BR 2497-2500. We 
report the hydrogen data for simplicity as those have the most statistical weight. However, the same conclusions are 
derived if we use the combination of hydrogen and positrons. 

\begin{table}[t]
    \begin{tabular}{ccccc}
         \hline
            $B_{\rm tot}(t)$ & $B_{\rm tot}(t)$ & $\alpha(t)$ & $\alpha(t)$ & $\Delta \chi^{2}$ \\
            time aver. \, & time delay \, & time aver. \, & time delay \, & from \\
            ($\#$BR) \, & ($\#$BR) \, & ($\#$BR) \, & ($\#$BR) \, & best fit  \\
            \hline \hline
            4  & 4 & 0 & 0 &  -- \\
            4 & 4 & 1 & 1 & 118  \\
            4  & 4 & 2 & 2 & 182 \\
            4  & 4 & 4 & 4 & 245 \\     
            9  & 9 & 0 & 0 & 1639 \\
            9  & 9 & 1 & 1 & 1734 \\
            6  & 6 & 0 & 0 & 1866 \\
            12 & 12 & 0 & 0 & 1912 \\
             1  & 1 & 0 & 0 & 1968 \\  
            \hline \hline 
        \end{tabular}
    \caption{A sample of time averaging schemes for the values of $B_{\rm tot}$ and $\alpha$, that the 
    positively charged particles experience. We give the corresponding $\Delta \chi^{2}$ from the best fit 
    choice relying on the \textit{AMS-02} hydrogen observations (see text for details).}
    \label{tab:Time_delay_fits}
\end{table}  

\begin{figure}
\vspace{-0.10in}
\hspace{-0.15in}
\includegraphics[width=3.72in,angle=0]{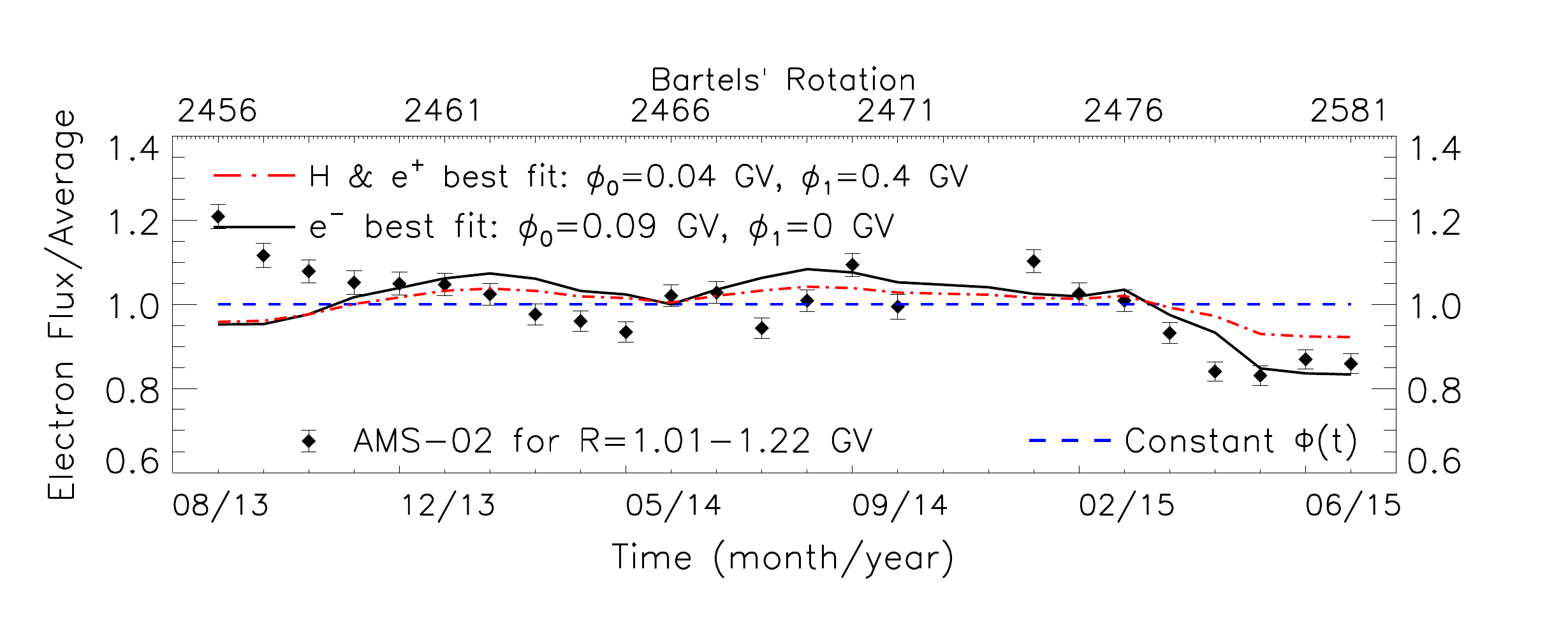}\\
\vspace{-0.15in}
\hspace{-0.15in}
\includegraphics[width=3.72in,angle=0]{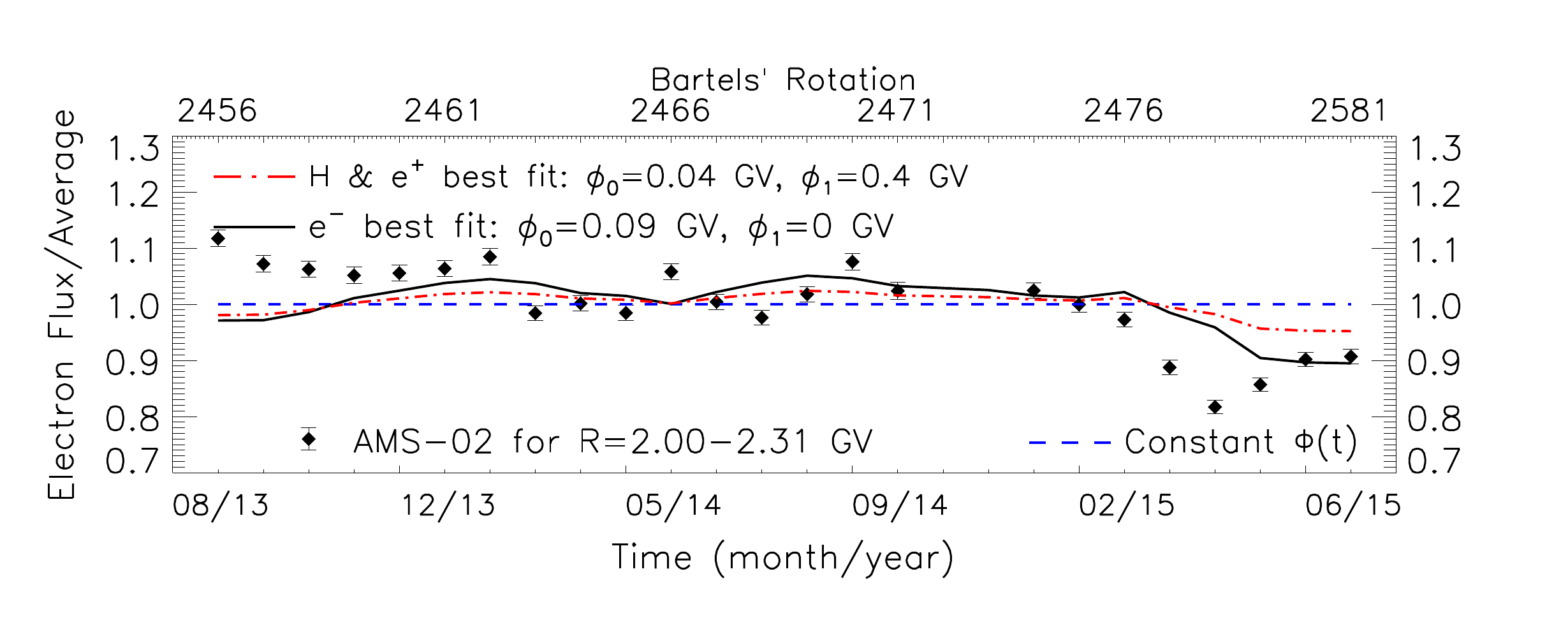}\\
\vspace{-0.15in}
\hspace{-0.15in}
\includegraphics[width=3.72in,angle=0]{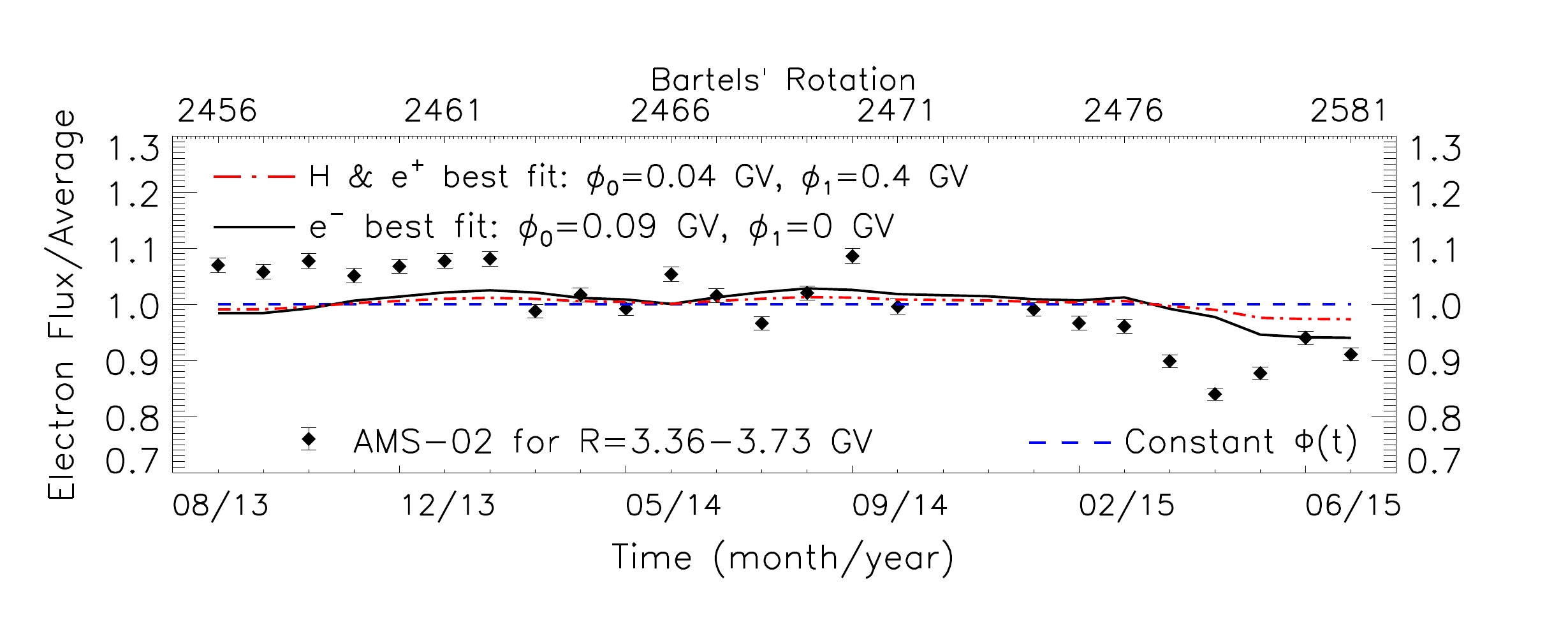}\\
\vspace{-0.15in}
\hspace{-0.15in}
\includegraphics[width=3.72in,angle=0]{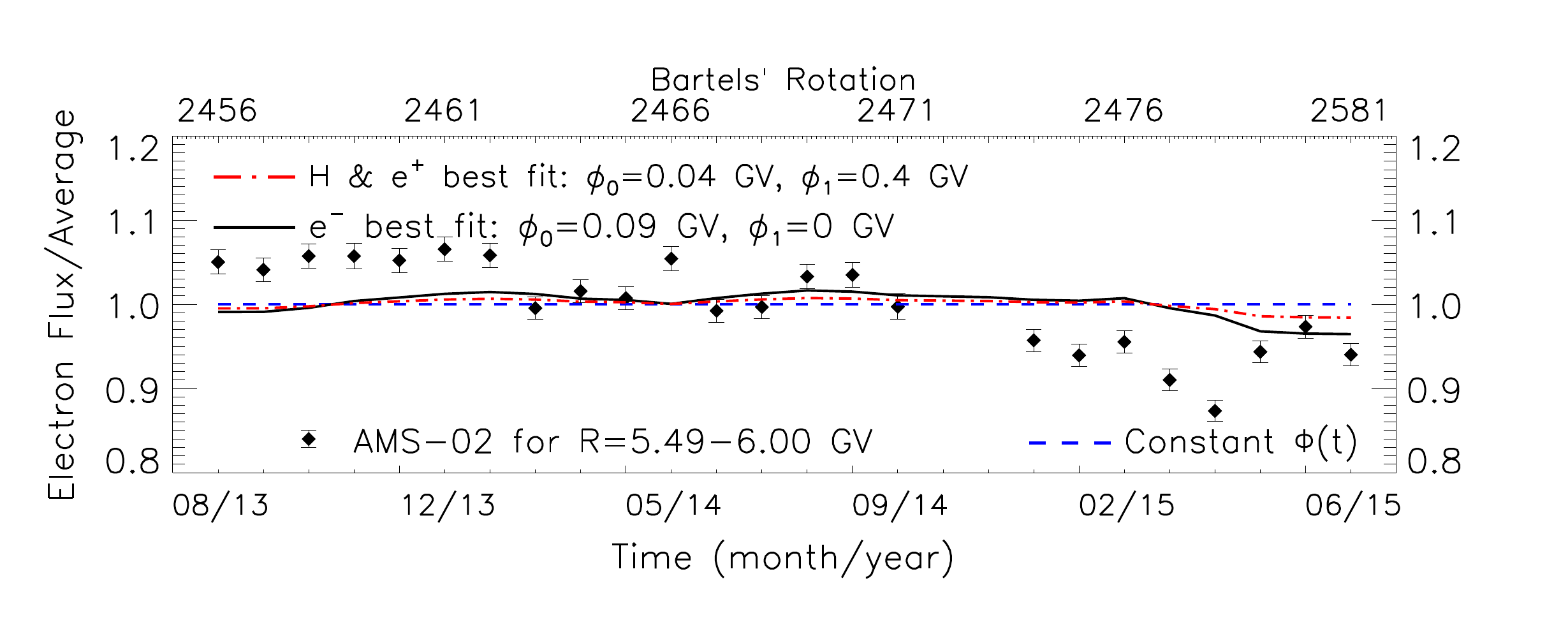}\\
\vspace{-0.15in}
\hspace{-0.15in}
\includegraphics[width=3.72in,angle=0]{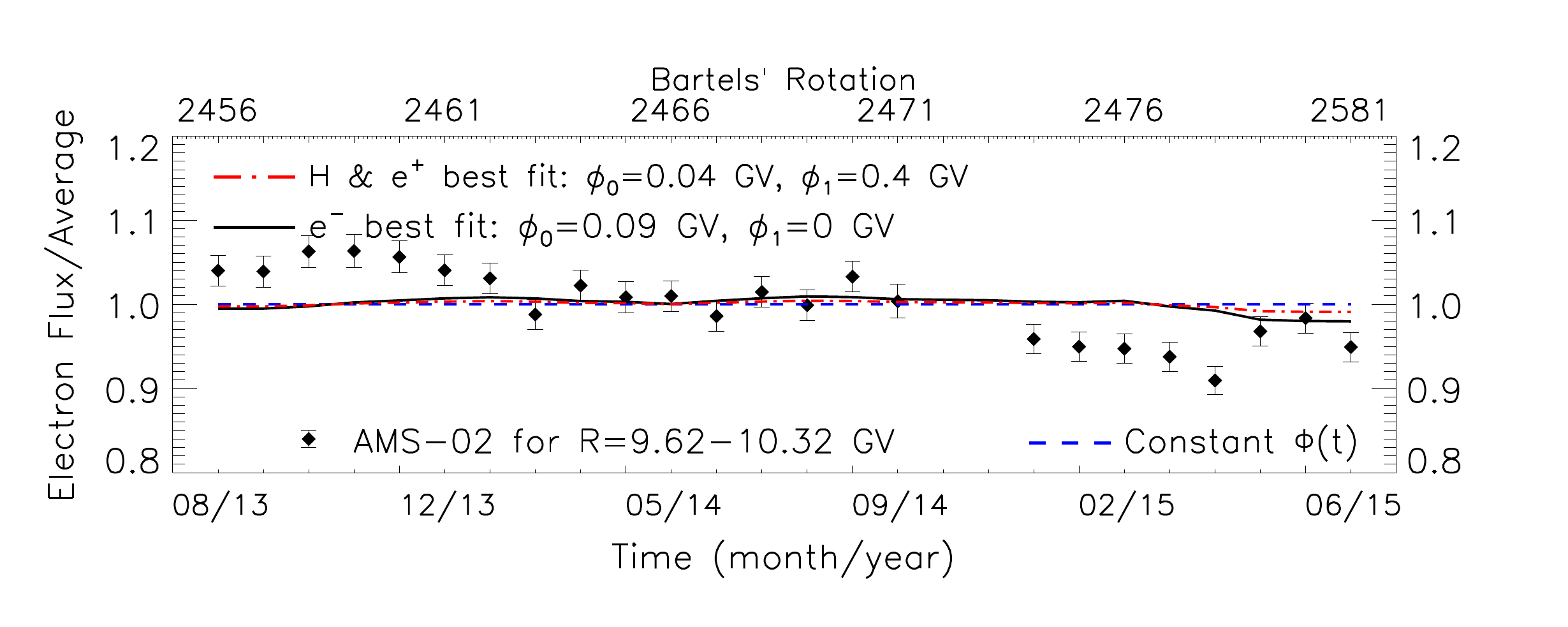}
\vspace{-0.20in}
\caption{Time evolution of the ratio of cosmic-ray electron flux to the averaged electron cosmic-ray flux 
within a period of Bartels' cycles 2456 and 2481. We show the same five rigidity bins as for positrons (see right 
column of Fig.~\ref{fig:PositiveCharges_BR2456-2481}), from top to bottom: 1.01-1.22, 2.00-2.31, 3.36-3.73, 
5.49-6.00, 9.62-10.32 GV. The solid black line gives the best fit choice for electrons derived for $\phi_{0} = 0.09$ 
GV, $\phi_{1} = 0$ GV and $R_{0} = 0.05$ GV (see text for details). The red dashed-dotted lines give the prediction for the 
evolution of the electron flux ratio using $\phi_{0} = 0.04$ GV and $\phi_{1} = 0.4$ GV that were derived to 
give the best fit to the combination of the positively charged cosmic-ray particles in the same era (see text for 
further details).}
\label{fig:Electrons_BR2456-2481}
\end{figure}

As we wrote, the hydrogen and positron cosmic-ray spectra constrain the values of $\phi_{1}$ and $R_{0}$ 
as well. We find that $\phi_{1} > 0$ is clearly preferred by the data even for the positively changed particles. 
This provides a clear indication for the charge-dependence of solar modulation on cosmic rays.  In 
Fig.~\ref{fig:PositiveCharges_BR2456-2481}, both the hydrogen and the positron flux ratios general time 
evolution for BR 2456-2481 can be described through a value of $\phi_{0} \sim 0.05$ GV and a $\phi_{1} \ge 0.3$.
The respective black lines provide the best fit to the five rigidity bins and 24 BR data points \footnote{In the fitting
process we included additional normalization coefficients for each of the rigidity bins. These are nuisance 
parameters that allow for the averaged ratios between different rigidity bins to further fluctuate. Our fits gave 
those parameters to be within 3$\%$ of unity.}. The red-dashed dotted lines instead assume that 
both hydrogen's and positrons' solar modulation is described by the same  combination of  $\phi_{0}, \;  \phi_{1}$
parameters. The best fit values for the combination of these species' flux evolution is $\phi_{0} =0.04$ GV and  
$\phi_{1}=0.4$ GV. The lines on the left and right are not identical due to the different masses of these particles. 
The time evolution of the hydrogen and positron fluxes are most prominent in the lowest two rigidity bins and 
present also in the bin around 3.5 GV. For rigidities above 5 GV while there are still statistically significant 
time-variations especially for the hydrogen flux, the time-patterns observed at low rigidities are not present.
In addition, our fitted model suggests very marginal time-evolution for the flux ratios at these higher rigidities.
In Fig.~\ref{fig:PositiveCharges_BR2456-2481}, we provide a blue dashed line that represents the effect of 
non-zero but also constant in time solar modulation potential $\Phi$.  

Our model does explain well the general time-evolution of solar modulation. This demonstrates that a simple 
analytic formula can explain the \textit{AMS-02} observations and connect the solar modulation of different 
cosmic-ray species to directly observable quantities of the HMF. However, statistically our best fits never approach
a $\chi^{2}$ per degree of freedom of approximately 1 that would suggest a proper good fit. At the higher rigidities
there are specific Bartels' cycles that deviate significantly from the observed averaged modulation (blue dashed lines),
and our model can not explain these short timescale observations. Such examples are BR 2457, BR 2463 and 
BR 2478. These are times that the incoming cosmic rays were more strongly affected by the HMF structure than 
our time-averaged assumptions predict; as we systematically under-predict the resulting total modulation potential 
value $\Phi$. There are however cycles like BR 2475 where at one rigidity bin the flux decreased compared to the 
neighboring in time Bartels' cycles and in the next rigidity bins it is increased. We believe that this indicates the 
level of stochasticity that solar modulation on the observed spectra is expected to have, and that is associated 
to the random paths of inwardly propagating cosmic-rays (see also \cite{Vittino:2019yme}).  

In Fig.~\ref{fig:Electrons_BR2456-2481}, we give the time evolution of the cosmic-ray electron flux ratio over 
BR 2456 to BR 2481. During that era the $\phi_{1}$ term was mostly shut off for electrons. Thus, the $\phi_{1} = 0$ 
and $R_{0} = 0.05$ best fit values while still statistically significant due to the small size of the \textit{AMS-02} error-bars 
do not necessarily exclude other ranges for the $\phi_{1}$ and $R_{0}$ combination as we will explain later. We note 
however, that the electrons observations are not well explained by either our best fit parameters (black lines) nor the 
choice of $\phi_{0} = 0.04$ GV and $\phi_{1} = 0.4$ GV that explain the combination of the positively charged 
particles in the same era (red dashed-dotted lines in Fig.~\ref{fig:Electrons_BR2456-2481}). In fact, the electron 
data of that era provide the greatest challenge of any species/era in 
our work. The issues we experience are beyond a small number of unique Bartels' rotation cycles.  Also, our model 
under-predicts the time evolution both at the lower and higher rigidity bins. Electrons seem to be the most affected 
by the polarity flip as with the $A<0$ before that flip they would have traveled to the Earth through the solar poles. 
Our model assumes that was still the case for most of the BR 2456 to BR 2481 era and the tension with the data,
suggests that their path through the Heliosphere during that transitional era was more complex.

\begin{figure*}
\begin{center}
\begin{tabular}{c c}
\vspace{-0.08in}
\hspace{-0.15in}
\includegraphics[width=3.72in,angle=0]{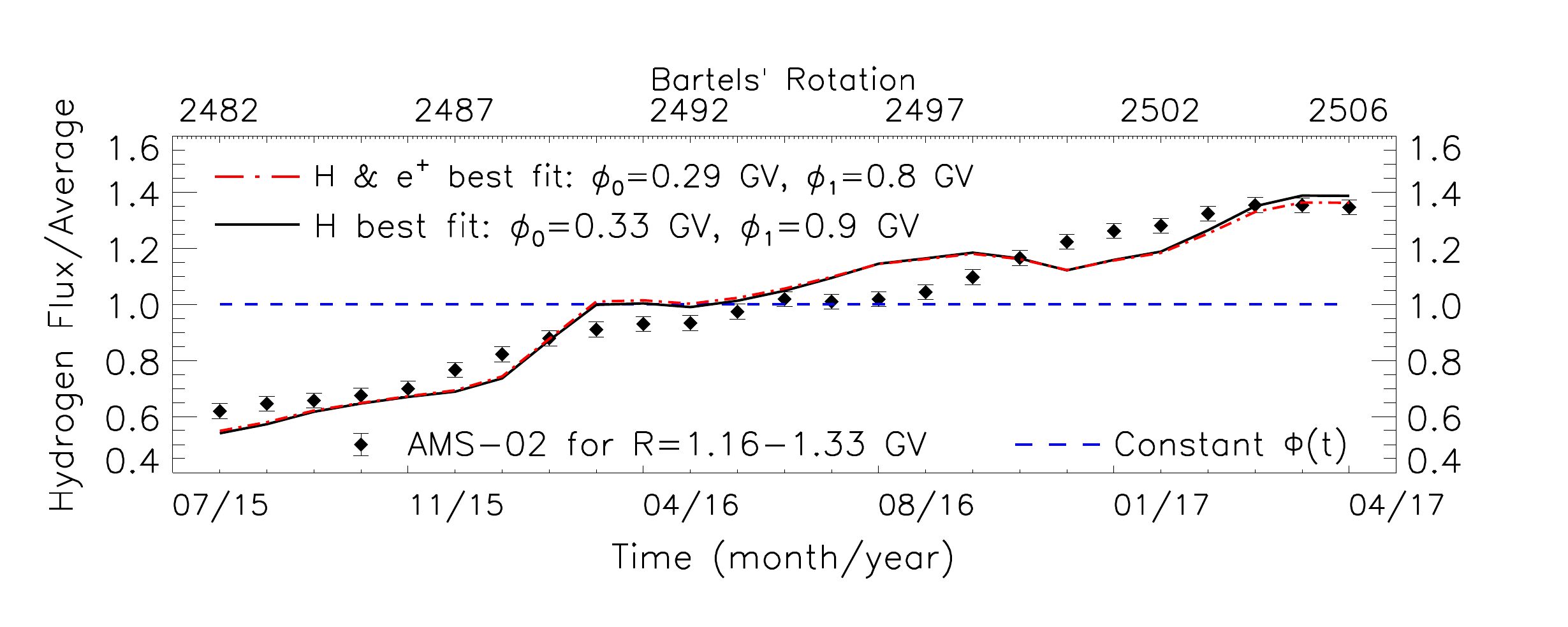}
\hspace{-0.15in}
\includegraphics[width=3.72in,angle=0]{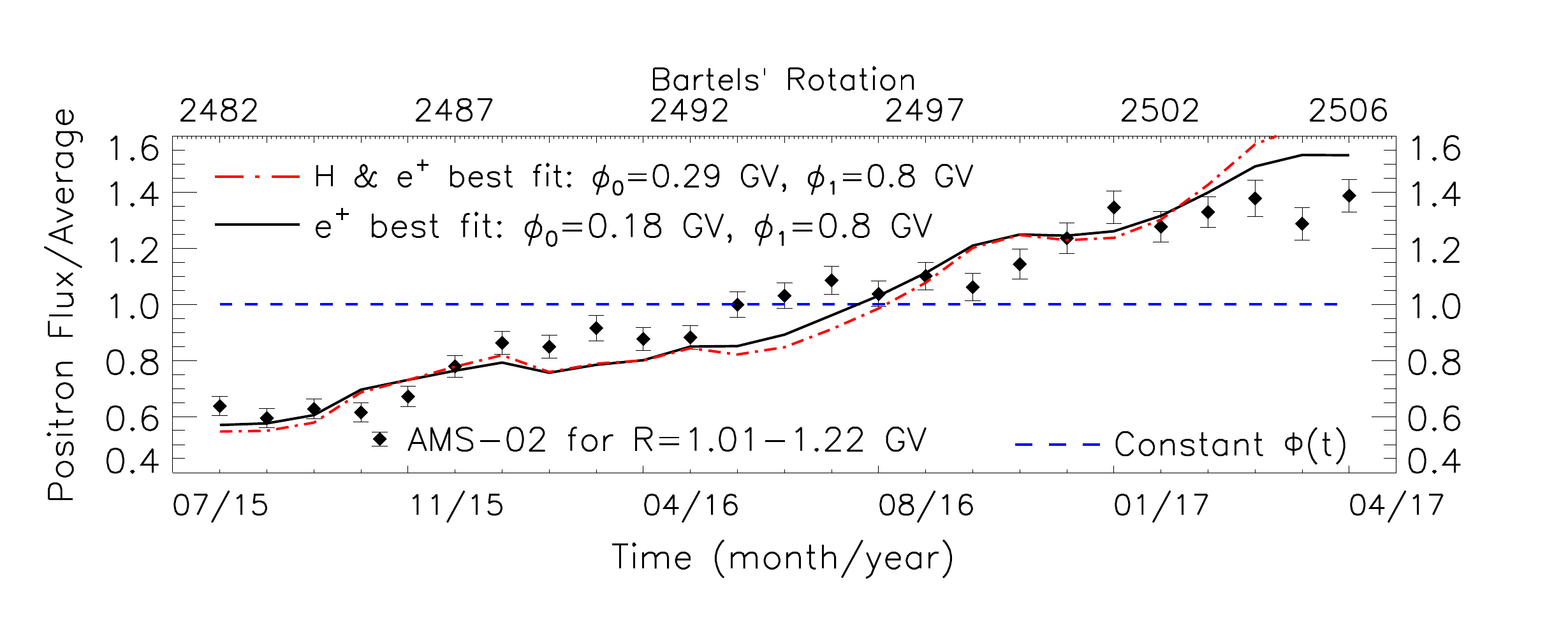}\\
\vspace{-0.08in}
\hspace{-0.15in}
\includegraphics[width=3.72in,angle=0]{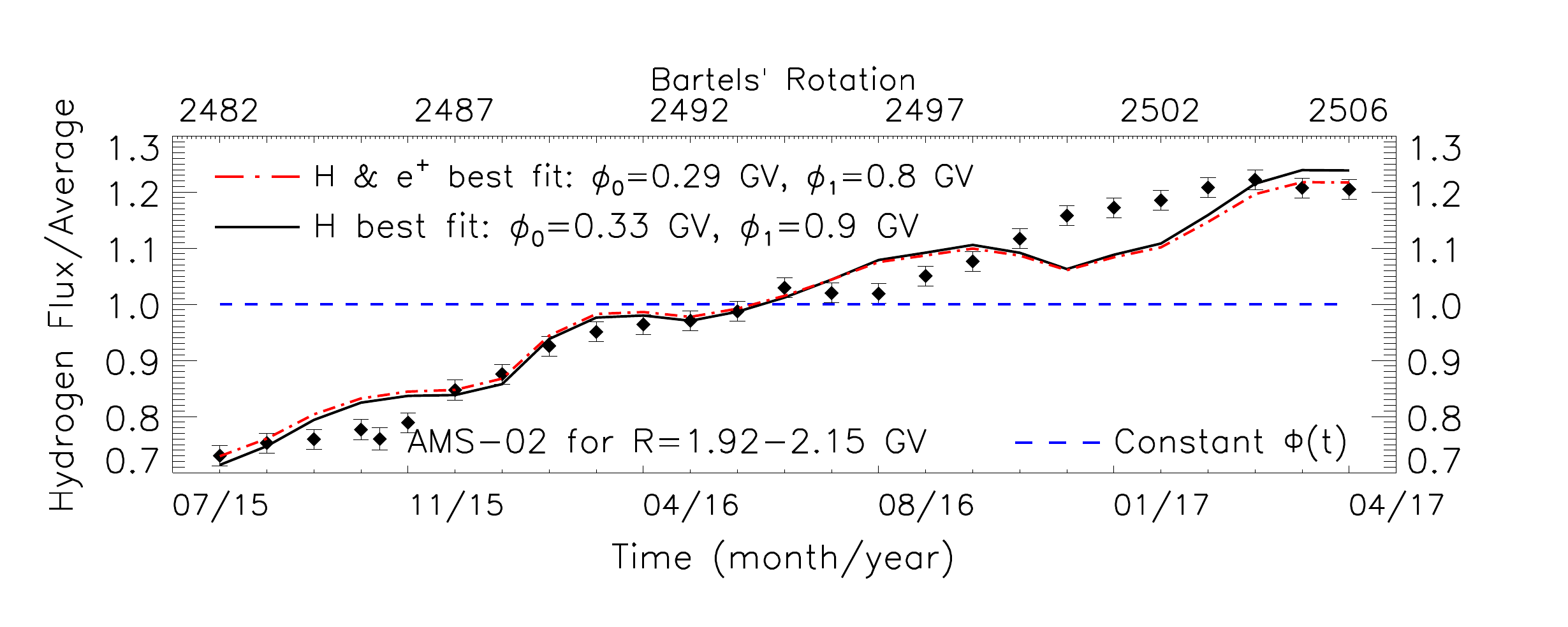}
\hspace{-0.15in}
\includegraphics[width=3.72in,angle=0]{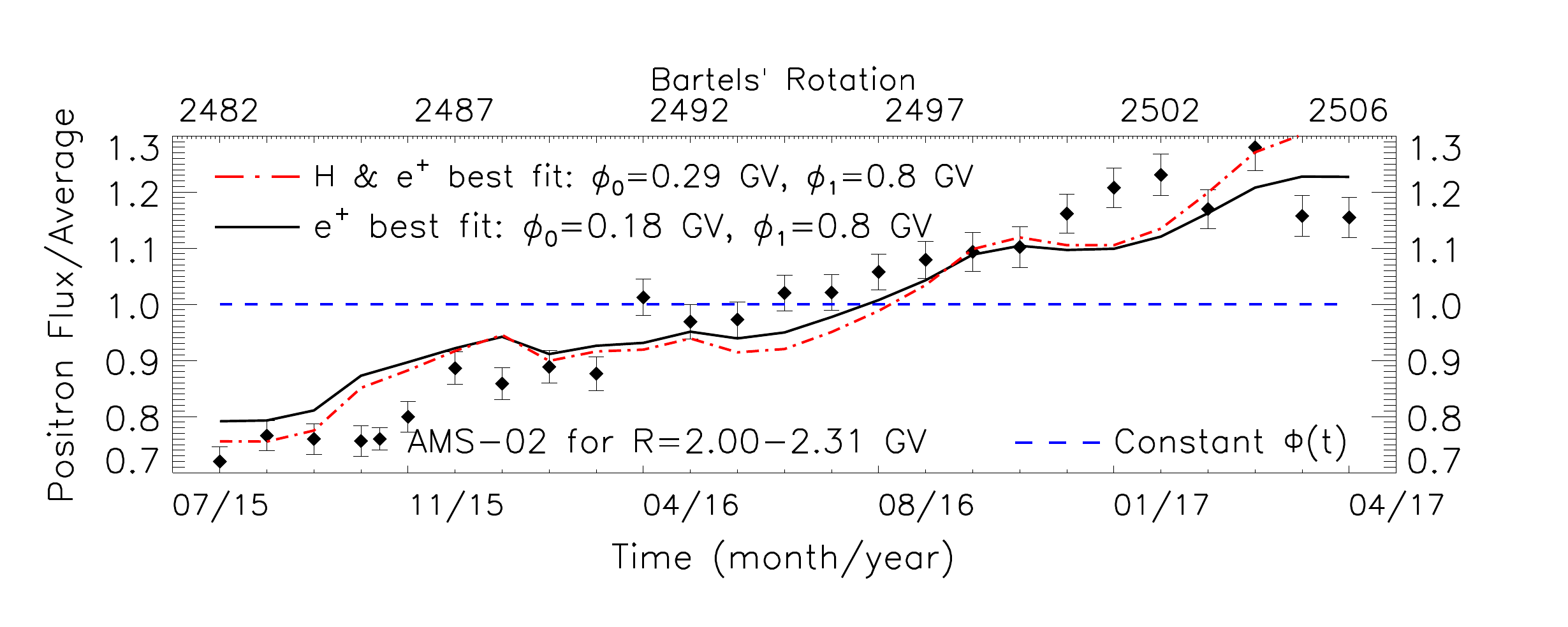}\\
\vspace{-0.08in}
\hspace{-0.15in}
\includegraphics[width=3.72in,angle=0]{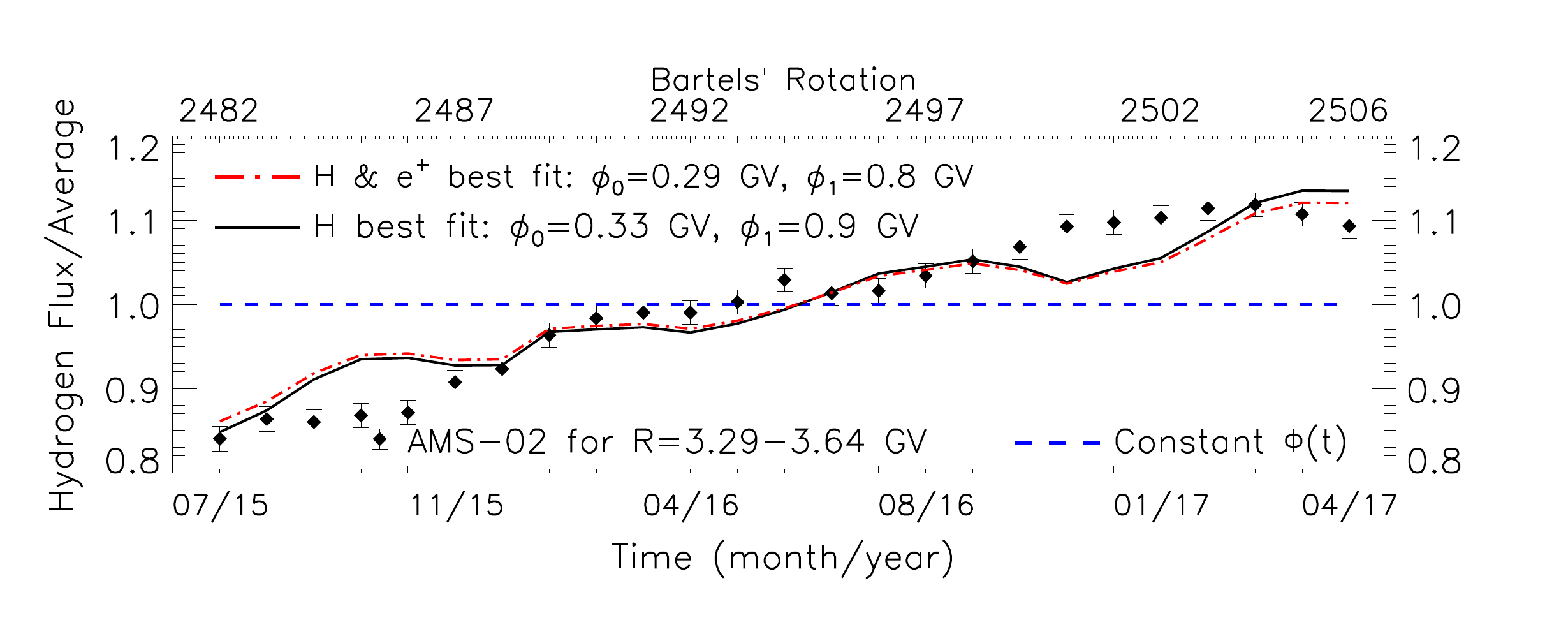}
\hspace{-0.15in}
\includegraphics[width=3.72in,angle=0]{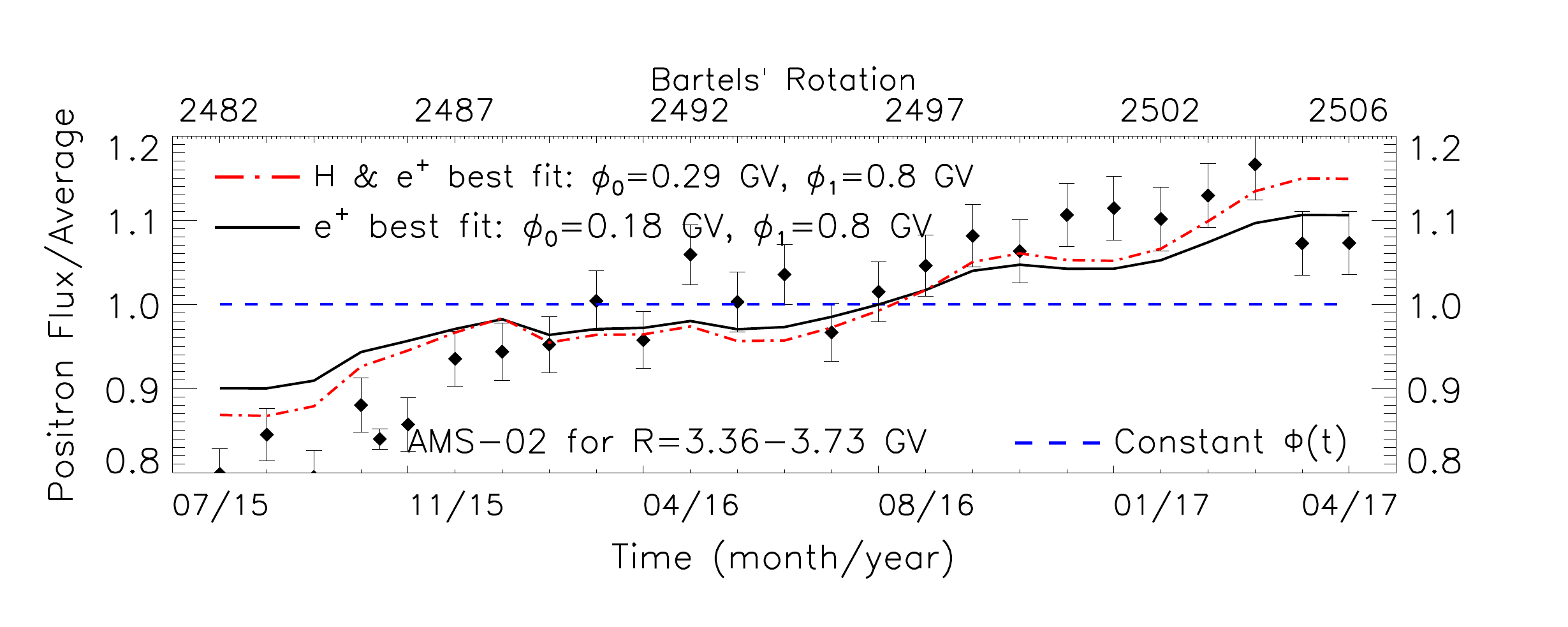}
\end{tabular}
\end{center}
\vspace{-0.15in}
\caption{As with Fig.~\ref{fig:PositiveCharges_BR2456-2481}, for the 25 observed Bartel's cycles of BR 
2482-2506. \textit{Left column}: the time evolution of the ratio of the cosmic-ray hydrogen flux to the 
averaged hydrogen cosmic-ray flux. We show the following three lower rigidity bins, from top to bottom: 
1.16-1.33, 1.92-2.15 and 3.29-3.64 GV. The black line gives the best fit choice for the hydrogen (H) ratio 
for $\phi_{0} = 0.33$ GV, $\phi_{1} = 0.9$ GV and assuming $R_{0} = 2$ GV. \textit{Right column}: the 
time evolution of the ratio of the cosmic-ray positron flux to the averaged positron cosmic-ray flux for 
the era of study. We show the positron rigidity bins of 1.01-1.22, 2.00-2.31 and 3.36-3.73 GV. The black line 
gives the best fit choice for the positron ($e^{+}$) ratio for $\phi_{0} = 0.18$ GV, $\phi_{1} = 0.8$ GV assuming 
$R_{0} = 1.0$ GV. The red dashed-dotted lines on both left and right columns give the evolution of the respective 
hydrogen and positron flux ratios as predicted for values of $\phi_{0} = 0.29$ GV and $\phi_{1} = 0.8$ GV, which
provide the best fit to the combination of both particle species at that era (see text for details).}
\label{fig:PositiveCharges_BR2482-2506}
\end{figure*}

In Fig.~\ref{fig:PositiveCharges_BR2482-2506}, we plot the time-evolution for the positively charged particles 
during the BR 2482 to BR 2506 era. During that time the $\phi_{1}$ term of Eq.~\ref{eq:ModPot}, gradually switched 
off for the positively charged particles, being present up to BR 2489. This choice is used to model the time it takes for positively
charged particles' trajectories to stabilize in reaching Earth through the Heliospheric poles. Any effect the 
\textit{AMS-02} data may have on $\phi_{1}$, $R_{0}$ parameters is solely dependent on the first eight (BR 2482-2489)
data points. In Table~\ref{tab:Phi1_switch_off}, we show the impact alternative options on the switching off of the $\phi_{1}$ term
have on the quality of the $\chi^{2}$-fit to the \textit{AMS-02} data. That switching off of the $\phi_{1}$ term, is
based on the best-fit averaging choices for $B_{\rm tot}(t)$ and $\alpha(t)$, that we discussed earlier. For 
that era these are also given in Table~\ref{tab:Time_delay_fits}.

From the BR 2482 to BR 2506 era, we find that fitting the hydrogen and positron 
flux ratio evolution independently or together has only a small effect on the derived modulation potential 
parameters. Hydrogen data require larger values of $\phi_{0} = 0.33$ GV and $\phi_{0} = 0.9$ GV compared to 
positrons, that give best fit for $\phi_{0} = 0.18$ GV and $\phi_{0} = 0.8$ GV. These are depicted by the respective
black lines on the left and right columns of Fig.~\ref{fig:PositiveCharges_BR2482-2506}. Our results rely on the same 
five rigidity bins used for  BR 2456 to BR 2481 of Fig.~\ref{fig:PositiveCharges_BR2456-2481}. However, as with 
that era the statistically most prominent rigidities are the lowest. The best-fit combination result for both positively 
charged cosmic-ray species is achieved for $\phi_{0} = 0.29$ GV and $\phi_{0} = 0.8$ GV, shown through the red
dashed-dotted lines. 

\begin{table}[t]
    \begin{tabular}{ccc}
         \hline
            Last BR with $\phi_{1}$-term & First BR with $\phi_{1}$-term & $\Delta \chi^{2}$ \\
            fully switched on \, & switched off \, & from \\
            ($\#$BR) \, & ($\#$BR) \, & best fit  \\
            \hline \hline
            2483 & 2490 &  -- \\
            2484 & 2491 & 21 \\
            2485 & 2491 & 40  \\ 
            2480 & 2503 & 60  \\
            2480 & 2501 & 63 \\
            2480 & 2500 & 76 \\ 
            2480 & 2499 & 88 \\
            2480 & 2497 & 116 \\
            2483 & 2494 & 162 \\    
        \hline \hline 
        \end{tabular}
    \caption{A variety of schemes for the switching off of the $\phi_{1}$ term of Eq.~\ref{eq:ModPot} with the corresponding 
    $\Delta \chi^{2}$ from the best fit option (see text for details).}
    \label{tab:Phi1_switch_off}
\end{table}           

\begin{figure}
\vspace{-0.10in}
\hspace{-0.15in}
\includegraphics[width=3.72in,angle=0]{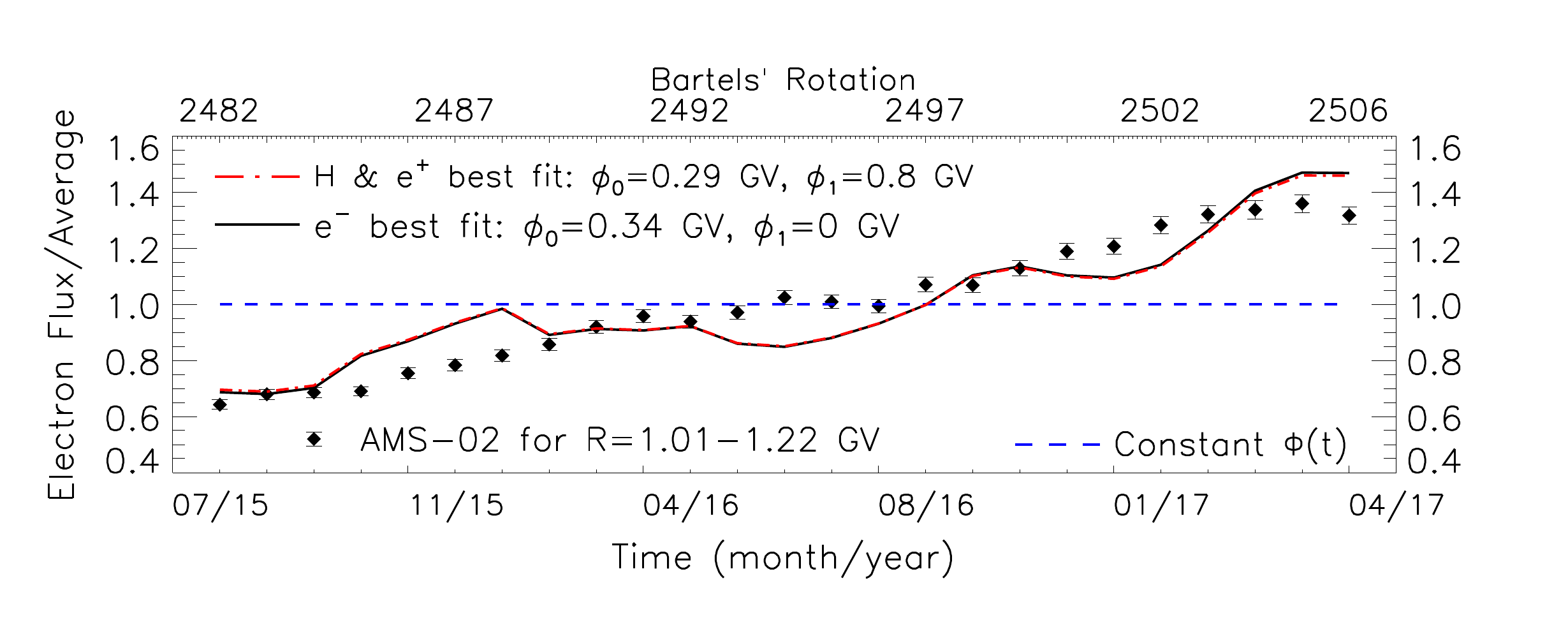}\\
\vspace{-0.15in}
\hspace{-0.15in}
\includegraphics[width=3.72in,angle=0]{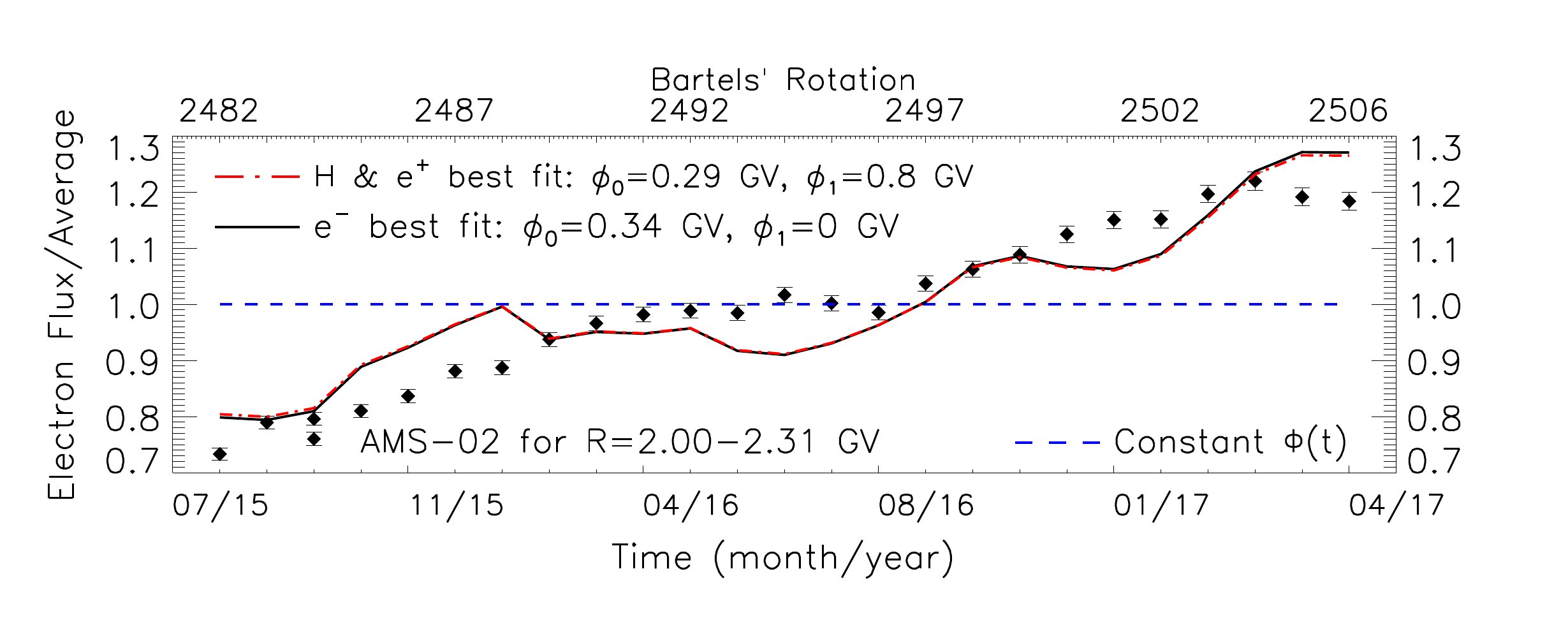}\\
\vspace{-0.15in}
\hspace{-0.15in}
\includegraphics[width=3.72in,angle=0]{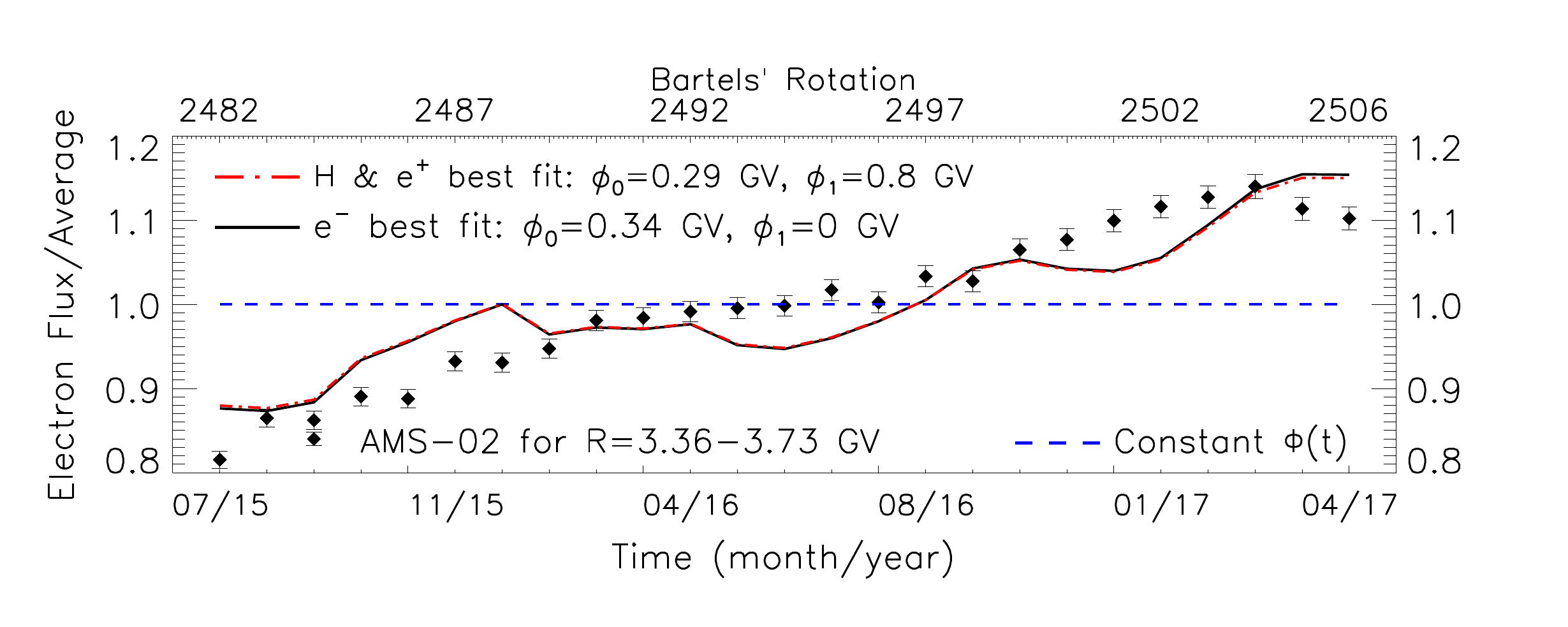}
\vspace{-0.20in}
\caption{As with Fig.~\ref{fig:Electrons_BR2456-2481}, for the 25 observed Bartel's cycles of BR 2482-2506.  
We show the first three rigidity bins as for electrons, from top to bottom: 1.01-1.22, 2.00-2.31 and 3.36-3.73 GV.
The solid black line gives the best fit choice for electrons derived for $\phi_{0} = 0.34$ GV, $\phi_{1} = 0$ GV 
and $R_{0} = 0.05$ GV. The red dashed-dotted lines give the prediction for the evolution of the electron flux ratio 
using $\phi_{0} = 0.29$ GV and $\phi_{1} = 0.8$ GV that give the best fit to the combination of the positively 
charged cosmic-ray particles in the same era (see text for further details).}
\label{fig:Electrons_BR2482-2506}
\end{figure}

In Fig.~\ref{fig:Electrons_BR2482-2506}, we show the evolution of the cosmic-ray electron flux between BR 2482 to 2506. 
Unlike the first era of study, our model describes the general trend of the electron flux increasing over time. We find that 
the best fit choice is achieved for a combination of $\phi_{0} = 0.34$ GV, $\phi_{1} = 0$ GV and 
$R_{0} = 0.05$ GV (black solid line). However, even the assumptions from fitting the positively charged particles of the
same era, i.e. $\phi_{0} = 0.29$ GV, $\phi_{1} = 0.8$ GV (red dashed-dotted lines), are in a similar level of agreement. 
In fact the $\phi_{1}$ parameter on electrons remains quite weakly constrained by the observations of that era. Yet,
the overall quality of fit still remains poor. The main reason for this tension is that our model gives a greater level of 
electron flux variability with time over the studied period than the \textit{AMS-02} data suggest. Other averaging effects 
may be in play, associated likely to the stochastic nature of paths followed by individual cosmic-rays before reaching us.
We leave that as a subject of future study, to be addressed with more observational data. 

\section{Conclusions and Discussion}
\label{sec:Conclusions}

\begin{figure*}
\begin{center}
\hspace{-0.35in}
\includegraphics[width=3.95in,angle=0]{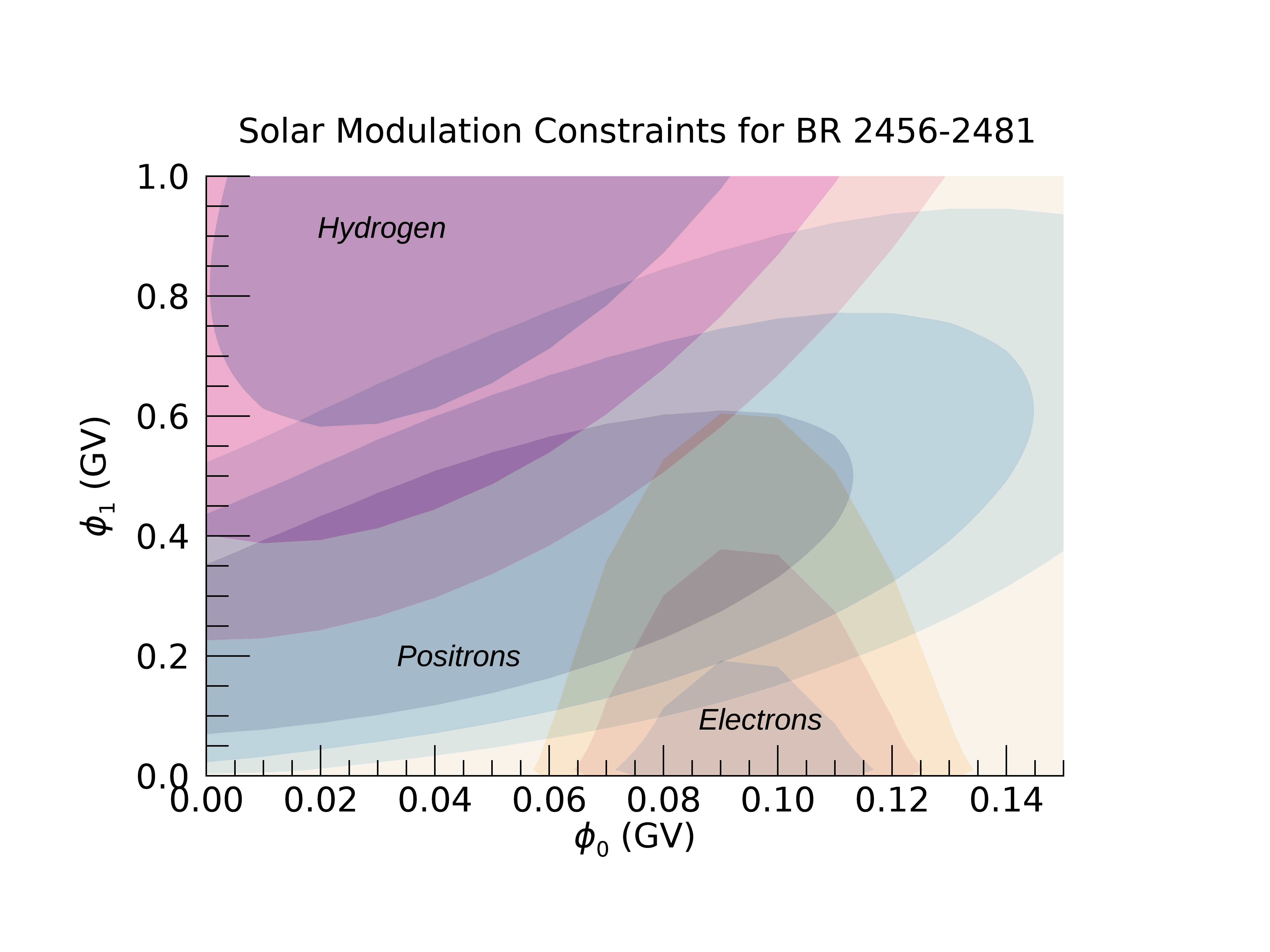}
\hspace{-0.70in}
\includegraphics[width=3.95in,angle=0]{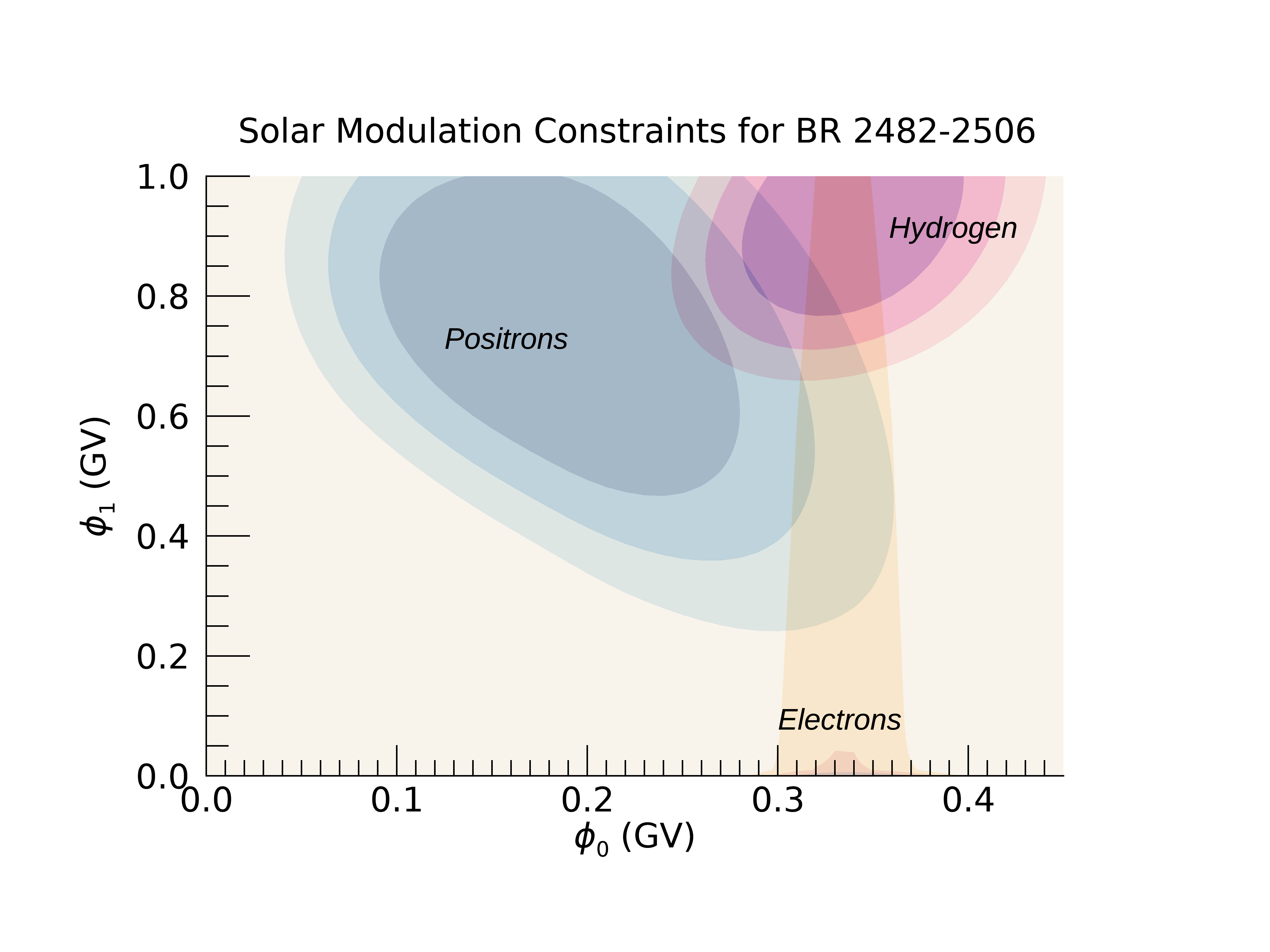}
\end{center}
\vspace{-0.45in}
\caption{The, 1-, 2- and 3-$\sigma$ best fit projected ranges for $\phi_{0}$ and $\phi_{1}$ parameters.
\textit{Left:} the allowed ranges from the era or Bartels' rotations 2456-2481 for hydrogen (purple), positrons (blue) and
electrons (orange). \textit{Right:} same as left but for the era or Bartels' rotations 2482-2506. There is a significant shift
in the $\phi_{0}$, $\phi_{1}$ parameter space between the two eras, that we associate to the fact that the heliospheric 
polarity that the cosmic rays experienced in the first of the two eras was not well defined (see text for further details).}
\label{fig:Parameter_Space_Ranges}
\end{figure*}

We conclude our work by presenting the $\phi_{0}$, $\phi_{1}$ projected space limits from all three types of cosmic 
rays; hydrogen (i.e. protons and deuterons), positrons and electrons. The 1-, 2- and 3-$\sigma$ acceptable ranges are
given in Fig.~\ref{fig:Parameter_Space_Ranges}. The hydrogen ratio ranges are provided with purple contours, while the
positron ratio ranges are given by the blue contours. Finally, the electron ratio ranges are given by the orange contours. 

In both eras there are combinations of $\phi_{0}$, $\phi_{1}$ values that can explain the observations from all three species,
i.e. $\phi_{0}$, $\phi_{1}$ where all three contours overlap (at the 3-$\sigma$-level). However, there are two major points 
to be made again. First in both cases, the electrons prefer a parameter space that is separeted from that of 
positively charged particles. That is especially evident on the $\phi_{1}$ values where electrons receive for both eras 
a best-fit value of $\phi_{1}=0$; whereas the positively charged species have a clear preference for $\phi_{1}>0$. 
For the first of the two eras our $\phi_{1}=0$ result, is quite anticipated as the electrons are expected to have 
mostly traveled through the poles and thus having had the $\phi_{1}$-term mostly turned off. For the second era of BR 
2482-2506, the $\phi_{1}=0$ best-fit result is less trivial to explain. Yet, the fit from that era does not constrain that parameter well. 
We also point out that the \textit{AMS-02} flux ratio data come with very small errors; thus any tension between the model
and the observations easily becomes statistically significant. Moreover, our model provides only a 
basic description of the overall patterns, but can not always explain the observed flux variations between successive 
Bartels' cycles as we have pointed out in the discussion around 
Figs.~\ref{fig:PositiveCharges_BR2456-2481},~\ref{fig:Electrons_BR2456-2481},~\ref{fig:PositiveCharges_BR2482-2506} 
and~\ref{fig:Electrons_BR2482-2506}.
The other major point is that the $\phi_{0}$, $\phi_{1}$  parameter ranges preferred between the two eras are distinctively 
different. The BR 2456-2481 prefers much smaller values of $\phi_{0} < 0.15$ GV, from all three species. Instead, from 
the BR 2482-2506 we get values of $0.05< \phi_{0} < 0.44$ GV, depending on the species used. That latter result is in 
much closer agreement with earlier analysis of the cosmic-ray observations from the solar cycle 23 performed in 
\cite{Cholis:2020tpi}, during which time the polarity that the incoming cosmic-rays experienced was very well defined.
From, that era's results we found a preference for  $\phi_{0}$ = 0.21-0.435 GV and $\phi_{1}$ = 1.15-1.95 GV. 

We believe that the era of BR 2456-2481 represents an example of what happens to cosmic rays that propagate inwards 
closely after a Heliospheric polarity flip. On one hand their travel time is increased. That point we have tested by comparing the 
required averaging schemes for $B_{\rm tot}(t)$ and $\alpha(t)$ used in Eq.~\ref{eq:ModPot} to describe the charge-, time-
and rigidity-dependence of solar modulation. We present the original publicly available \textit{ACE} and WSO data in
Fig.~\ref{fig:Tilt_BField} and the averaged ones in Fig.~\ref{fig:AveragedHMF}. At the same time the actual energy losses
the cosmic-rays experience are smaller than during times of established polarity as was the BR 2482-2506 era or the 
solar cycle 23 \textit{AMS-02} observations. This suggest that cosmic-rays during the era of non-well established polarity 
of within the volume of the Heliosphere (not just the surface of the Sun) travel through parts of it with weak magnetic fields. 

We also point out that while the presence of the solar wind's non-zero bulk speed is crucial in explaining the presence of 
drift effects along the HCS, its time-evolution is quite minimal and does not explain the observed time-variations of the 
measured cosmic-ray fluxes. This statement is true  at least to first order level. Instead, our model does successfully 
associate the measured time evolution of $B_{\rm tot}(t)$ and $\alpha(t)$ to the observed cosmic-ray fluxes time-evolutions 
given in Figs.~\ref{fig:PositiveCharges_BR2456-2481}, ~\ref{fig:Electrons_BR2456-2481}, for the era of BR 2456-2481 
and Figs.~\ref{fig:PositiveCharges_BR2482-2506} and~\ref{fig:Electrons_BR2482-2506} for the BR 2482-2506. 

The assumptions we make in this paper while still rely on the basic assumption that solar modulation of cosmic-ray fluxes 
can be described though a shift in the averaged kinetic energy of cosmic rays given by the modulation potential $\Phi$, 
that modulation potential's value can be associated to the observable properties of the Heliosphere, which in our model 
following Ref.~\cite{Cholis:2015gna}, are the measured amplitude of the magnetic field $B_{\rm tot}(t)$ at 1AU and HCS 
title angle $\alpha(t)$. Our $\phi_{1}$ term in Eq.~\ref{eq:ModPot}, separates particles of opposite charge to account for 
the fact that depending on the polarity of the Heliospheric magnetic field cosmic rays travel on average through different 
volumes of the Heliosphere. Moreover, that term accounts for the presence of drift effects from the solar wind on the 
incoming cosmic rays that propagate through the HCS.  Thus, our model explicitly breaks away from the conventional 
1-D approach of the force field approximation \cite{1968ApJ...154.1011G} and while analytic it follows more closely to 
the lessons learned from more recent 3-D simulation work done by several authors as in~\cite{1983ApJ...265..573K, 
2004JGRA..109.1101C, Potgieter:2013pdj,POTGIETER2017848, 2004JGRA..109.1101C, 2012Ap&SS.339..223S, 
Qin2017, 2018ApJ...858...61B, 2019arXiv190307501J, Caballero_Lopez_2019, 2019ApJ...878...59B}. Furthermore, we 
have used the time-rich cosmic-ray and heliospheric magnetic field observations, to make a connection between the 
conditions in the Heliosphere and its effect on the measured cosmic-ray spectra. Our work does not aim to replace 
the achievements  performed on the numerical side but provides a simple formula to account for solar modulation 
that through the continuous observations becomes better constrained. That in turn allows us to include in a computationally 
efficient way the effects of solar modulation when studying other types of astroparticle questions as the sources 
and environmental conditions of cosmic rays, their propagation through the ISM and in turn the ISM conditions. 
Another benefit of implementing a better constrained analytic prescription for the solar modulation, is the search for 
exotic sources of cosmic-rays as dark matter; that could contribute to the antimatter particle fluxes. 

As future improvements, we need to account for the fact that the observed cosmic-ray fluxes are the result of particles
having reached our detectors in a stochastic manner. Thus, while we assume averaged properties for the $B_{\rm tot}(t)$ 
and $\alpha(t)$ time-dependent quantities that we use to model the averaged energy losses expected in solar modulation, we need to account for the fact that 
cosmic-ray particles of a given species will have a scatter in these energy losses. We believe that may explain -especially
for the case of the cosmic-ray electrons- some of the observed tensions between our model's expectations and the 
observations. Another major improvement to be pursued for future work, is to understand better how the solar wind and its 
embedded magnetic field change with time at different radii. Here we rely on the timed measurements on a single radial 
position (at 1 AU). A model that can include the time evolution of the solar wind at different radii can be of great benefit in that goal. 
An example of work in that direction has been recently pursed in \cite{Li:2022zio}. Furthermore, in addition to the continuous observations by \textit{AMS-02},
\textit{ACE} and WSO, measurements from the \textit{Parker Solar Probe} \cite{2016SSRv..204....7F} and the \textit{Solar 
Orbiter} \cite{2020A&A...642A...1M} are going to provide useful future insights on the properties of the inner Heliosphere. 

 \textit{Acknowledgements:} 
We would like to thank Tim Linden and Dan Hooper for collaboration on related work and in early stages of this project. We also acknowledge the use of \texttt{GALPROP} 
\cite{GALPROPSite, galprop, NEWGALPROP}.
IC acknowledges support from the Michigan Space Grant Consortium, NASA Grants No. NNX15AJ20H and No. 80NSSC20M0124.
IC acknowledges that this material is based upon work supported by the U.S. Department of Energy, Office of Science, Office of High Energy Physics, under Award No. DE-SC0022352.

\bibliography{PaperDraft}

\end{document}